\documentclass{eusflat2021}
\usepackage{cite}
\usepackage{amsmath,amssymb,amsfonts,lipsum,float,gensymb,cancel,mathrsfs,url,cancel,amsfonts,textcomp,xcolor,wrapfig,sidecap,pdfpages,fancyhdr,booktabs,multirow,capt-of,tabularx,changepage,transparent,etoolbox,color,amsmath,soul,lipsum,bm,rotating,caption,subcaption,stfloats}
\usepackage{algorithmic}
\usepackage{graphicx}
\usepackage{textcomp}
\usepackage[export]{adjustbox}

\title{\bf Simulating single-photon detector array sensors for depth imaging}
\author{Stirling Scholes$^1$, Germ\'an Mora-Mart\'in$^2$, Feng Zhu$^1$, Istvan~Gyongy$^2$, Phil Soan$^3$, and, Jonathan Leach$^{1*}$.\\
$^1$School of Engineering and Physical Sciences, Heriot-Watt University, Edinburgh, EH14 4AS, UK\\
$^2$School of Engineering, The University of Edinburgh, Edinburgh, EH9 3FF, UK \\
$^3$Cyber and IS Division, Defence science and technology laboratory, Porton Down, SP4 0JQ, UK\\
$^*$\email{j.leach@hw.ac.uk}}

\begin{document}

\maketitle

\begin{abstract}
Single-Photon Avalanche Detector (SPAD) arrays are a rapidly emerging technology. These multi-pixel sensors have single-photon sensitivities and pico-second temporal resolutions thus they can rapidly generate depth images with millimeter precision. Such sensors are a key enabling technology for future autonomous systems as they provide guidance and situational awareness. However, to fully exploit the capabilities of SPAD array sensors, it is crucial to establish the quality of depth images they are able to generate in a wide range of scenarios. Given a particular optical system and a finite image acquisition time, what is the best-case depth resolution and what are realistic images generated by SPAD arrays? In this work, we establish a robust yet simple numerical procedure that rapidly establishes the fundamental limits to depth imaging with SPAD arrays under real world conditions. Our approach accurately generates realistic depth images in a wide range of scenarios, allowing the performance of an optical depth imaging system to be established without the need for costly and laborious field testing. This procedure has applications in object detection and tracking for autonomous systems and could be easily extended to systems for underwater imaging or for imaging around corners.
\end{abstract}

\section{Introduction}
\label{sec: Introduction}
Light detection and ranging (Lidar) enabled by Single-Photon Avalanche Detector (SPAD) sensors has seen significant research in recent years as a next generation imaging technology~\cite{bellisai2010single,villa2014cmos,villa2021spads}. By coupling Silicon Semiconductor pixels to Time-to-Digital Converters (TDCs) imaging systems with both single-photon sensitivity and picosecond temporal resolution can be created, allowing for, in principle, the measurement of an objects `depth' (i.e., its surface profile) at millimeter scales~\cite{morrison202064,gramuglia2021low}. The ability to measure depth has led to SPADs being used in a wide range of applications, for instance: exploiting the single-photon sensitivity to image underwater~\cite{maccarone2015underwater}, through obscurants~\cite{tobin2018depth,tobin2019three} or at long range~\cite{mccarthy2009long,li2021single}; or, leveraging temporal gating to image `through' nets~\cite{tachella2019real,mau2020through} or to view the flight of a laser pulse~\cite{gariepy2015single,morland2021intensity}. More recently, SPAD-based Lidars have been examined for automotive applications~\cite{niclass20130,seo202036,ito2017spad} as well as for machine vision tasks including gesture recognition~\cite{gyongy2021high,hutchings2019reconfigurable}, identification~\cite{caramazza2018neural}, drone tracking~\cite{scholes2022dronesense}, and pose detection~\cite{ruget2021real}.\\
\\
In conjunction with the adoption of SPAD based imaging systems several publications have focused on modeling the performance of SPADs. These works are predominantly focused on two areas, either, simulating the electron avalanche characteristics of individual pixels, or, examining methods for better data processing. In the case of the former, analyses of the peak detection efficiency, timing resolution, jitter, and dark count rates have been achieved using physical models~\cite{gulinatti2009design,gulinatti2009modeling,panglosse2021modeling}, software~\cite{xu2016new,panglosse2020dark,poushi2020comprehensive}, Monte-Carlo simulations~\cite{lu2019monte}, and analytical models~\cite{sun2019simple}. Further, the effect of specific factors such as tunneling~\cite{cheng2016comprehensive}, nano-structures~\cite{ma2015simulation}, and temperature~\cite{shin2017structure} have also been examined.\\
\\
As well as examinations of the electron avalanche characteristics several works have focused on the processing of SPAD data. To create a depth image the timing information from multiple laser pulses is accumulated into a histogram. The position of the peak of the histogram is assumed to correspond to the depth of the target. Consequently the quality of the depth image produced by a SPAD imaging system is directly dependent on the ability to identify the peak of the histogram. This process is complicated by spurious counts in the histogram resulting from dark counts, solar background counts, or, back reflections from scattering media~\cite{gyongy2021direct}. Prior works have addressed this by adjusting TDC triggering based on adaptive photon thresholds~\cite{incoronato2022single,padmanabhan2019modeling,beer20182}, laser pulse modulation~\cite{ding2022coded,tsai2020anti,carrara2019optical}, attenuation~\cite{gupta2019photon}, principle component analysis for rapid processing~\cite{duan2022pca}, and match filtering~\cite{meng2021spad,sang2020mitigating,mau2020use}.\\
\begin{figure}[t!]
\centering
    %\begin{adjustbox}{max width=\columnwidth,center}
	\includegraphics[width=\columnwidth]{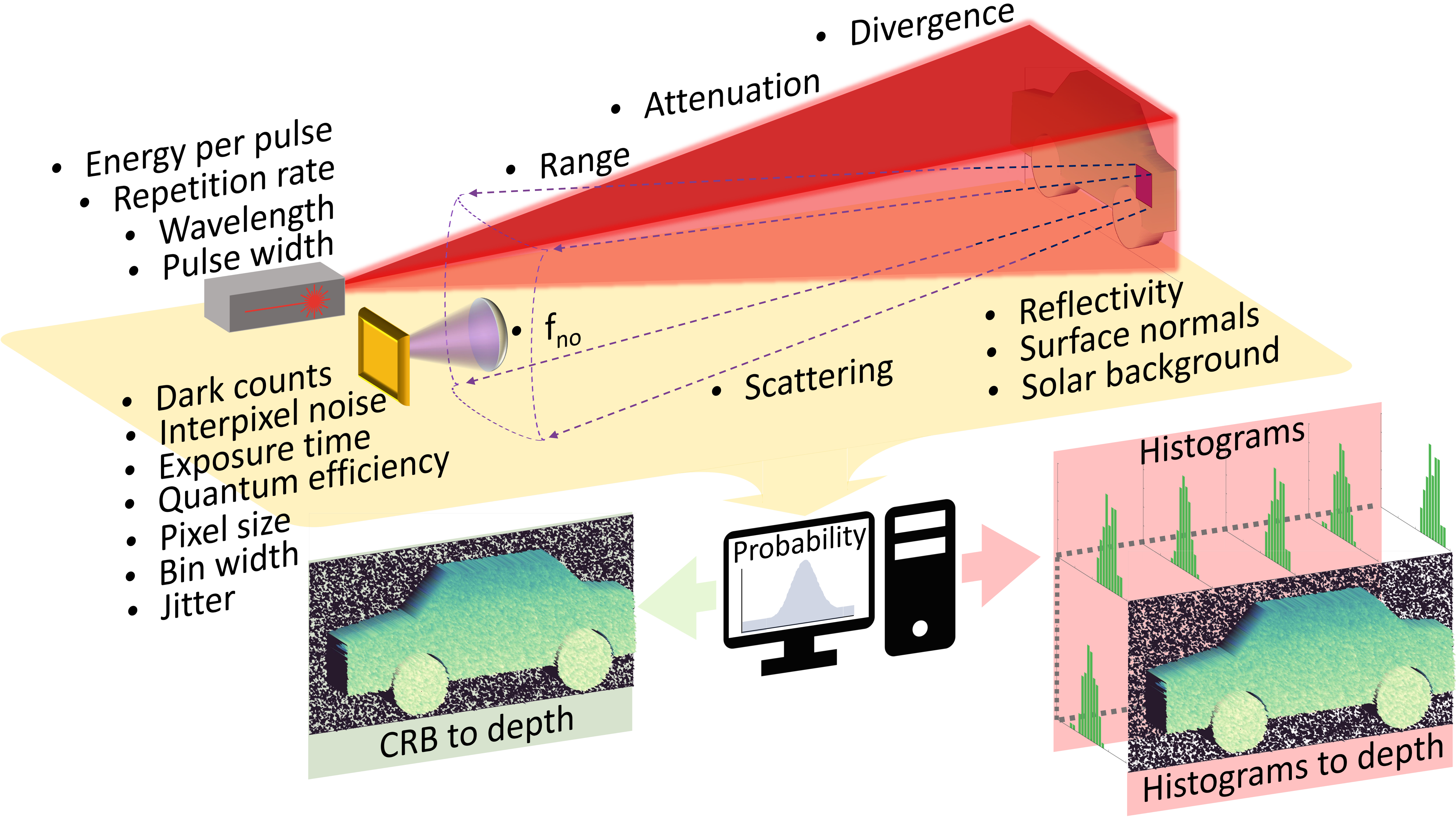}
        \caption{A conceptual summary of this work. In a direct time-of-flight flash Lidar system a laser pulse is used to illuminate a target as depicted by the red beam and simple vehicle model. Each pixel in the SPAD sensor then recieves light from a region of the target plane (illustrated by the square on the vehicle) subject to the properties of the target, scattering, the aperture of the collection lens, and the parameters of each pixel. The 19 bulleted parameters are used to define a computational model for photon measurements in the context of probabilities and Fisher information. This model can be used in two regimes: A computationally efficient (but inflexible) mode which directly simulates depth images by leveraging the Cram\'er-Rao bound (CRB); A computationally intensive mode which is highly flexible. This second mode precisely mimics SPAD operation by producing physically realistic histograms (and by extension depth images) on a per-pixel basis.}
	    \label{fig: Concept_Fig}
\end{figure}\\
Despite the growing number of SPAD-based Lidar applications, comparatively few works focus on modeling the performance of such systems in the context of entire images. Specifically, uncertainty in depth estimation has been examined using Monte-Carlo simulations and Fisher information for single pixels in the case of both simple and complex surfaces~\cite{koerner2021models,tontini2020numerical}. Further, semi-analytical single pixel models that characterise the SPADs using Poisson or Erlang distributions together with confidence intervals~\cite{beer2017range,beer2017modelling} and incorporate effects such as SPAD dead time~\cite{arvani2018direct,beer2017expected} have been presented. However, the Poisson (and by extension Erlang) distributions have been found to diverge from observed SPAD behaviour in both the small and large photon number limits~\cite{donati2020analysis,houwink2021theoretical}. Additionally, accurate simulations of SPAD images have been presented in the context of Fluorescence Lifetime Imaging Microscopy (FLIM) and lifetime estimation~\cite{houwink2021theoretical,wang2015modeling}.\\
\\
In this work we build upon the existing examinations of SPAD Lidar systems by combining a physical model with a binomial sampling procedure and a 3-Dimensional (3D) virtual environment to extend the accurate estimation of SPAD Lidar capabilities to entire images. A summary of this work is depicted conceptually in Fig.~\ref{fig: Concept_Fig}. Specifically, by accounting for the salient factors of real imaging systems, as bulleted in Fig.~\ref{fig: Concept_Fig}, we derive a physical process which directly relates SPAD measurements to the parameters of the optical system. By integrating this process with a 3D virtual environment (realised by the Unreal Engine) we develop a description of SPAD imaging in the context of photon arrival probabilities and Fisher information. The use of a 3D virtual environment allows for the application of our model to a wide variety of SPAD imaging scenarios since the complex geometries of real world objects, for instance the car depicted in Fig.~\ref{fig: Concept_Fig}, can be accurately accounted for. We develop a fast computational model for SPAD images by implementing our calculation of photon arrivals on a Graphics Processing Unit (GPU). We make this model publicly available at (https://github.com/HWQuantum/Simulating-single-photon-detector-array-sensors-for-depth-imaging).\\ 
\\
The computational model is capable of operating in two regimes; a computationally intensive but highly flexible and precise mode, or, a computationally efficient mode subject to some restrictions. In the one regime (the PC's right hand side output in Fig.~\ref{fig: Concept_Fig}) we employ a binomial sampling process to precisely mimic the operation of SPAD sensors. This process allows for the generation of physically realistic SPAD histograms (and by extension depth images) on a per pixel basis over an entire sensor. In the other regime (the PC's left hand side output in Fig.~\ref{fig: Concept_Fig}), we demonstrate that (under appropriate conditions) the Cram\'er-Rao bound associated with the Fisher information of the SPAD system can be used to model SPAD images directly. This direct modeling removes the need for histogram generation significantly reducing computational complexity. We confirm the validity of our approaches with experimental measurements of a resolution test target using a state-of-the-art SPAD array sensor and present median simulation accuracies in excess of $80\%$. Finally, we highlight the flexibility of our system by accurately modeling the SPAD images obtained of a vehicle at 1.4 km under real world conditions.
\section{Theory}
\label{sec: Theory}
The number of photons returned from a target is described by a photon channel that models the loss of signal photons as a series of sequential processes. Consider a laser pulse of wavelength $\lambda$ and initial energy $E_0$ having a divergence $\theta$ projected over a range $R$, through an atmosphere of attenuation length $C_{atm}$ as depicted in Fig.~\ref{fig: Concept_Fig}. For a target with Lambertian scattering and reflectivity $\Gamma$, the average number of photons detected per-pulse-per-pixel $P_{pp}$ by an imaging sensor with pixels of quantum efficiency $q$ and effective (pixel dimension $\times$ fill factor) size $W_p\times H_p$ at the focal plane of a collecting lens with f-number $f_{no}$ is
\begin{equation}
\begin{aligned}
    P_{pp} = \frac{\lambda E_0}{h c}\frac{q\Gamma e^{\frac{-2R}{C_{atm}}}}{8}\frac{ W_pH_p}{f_{no}^2\pi R^2\tan^2(\theta)}.
\end{aligned}
\label{eqn: photons per pulse}
\end{equation}
Here, $h$ is Plank's constant and $c$ is the speed of light. The $8$ in the denominator is a consequence of treating the surface reflectivity $\Gamma$ and the atmospheric scattering $C_{atm}$ of the return as two separate processes. If a different scattering mechanism is examined, the value in the denominator will change. Additional information on the derivation of Eq.~\ref{eqn: photons per pulse} is given in Sec.~\ref{subsec:Derivation of  photons detected per-pulse-per-pixel}. Note that even if $P_{pp}>1$ then only a single photon is measured due to the single-photon-per-pulse limit for SPADs. For many SPAD imaging applications the detected number of photons-per-pulse-per-pixel is less than one hence, Eq.~\ref{eqn: photons per pulse} represents the probability of a signal photon being measured by the detector. i.e., the SPAD can be expected to click at least every $1/P_{pp}$ pulses on average. Further, for clarity of explanation, we assume only a single wavelength for Eq.~\ref{eqn: photons per pulse}. In the most general case Eq.~\ref{eqn: photons per pulse} could be extended to multiple wavelengths by integrating over all relevant $\lambda$'s. Equation~\ref{eqn: photons per pulse} is a variant of the radar equation with $\Gamma$ functioning as an optical cross-section and $( W_pH_p)/[f_{no}^2\pi R^2\tan^2(\theta)]$ defining the relationship between the effective aperture and the illuminated area. This ratio indicates that larger pixels are desirable since in the limit of a single-point detector that sees the whole field-of-view all of the gathered energy is collected by a single pixel. Conversely, for a fixed sensor footprint increasing the transverse resolution of the detector will reduce the light gathering capabilities of each pixel. Note that Eq.~\ref{eqn: photons per pulse} is independent of the focal length of the collecting lens. This is because as the range increases, the focal length of the collecting lens must also increase to maintain the field-of-view. If the f-number is constant, this corresponds to a lens with a larger aperture which captures more of the scattered light and offsets the scattering losses. 
\subsection{Fisher information}
\label{subsec: Fisher info}
\begin{figure*}[b!]
\centering
    %\begin{adjustbox}{max width=\columnwidth,center}
	%\includegraphics[width=0.95\columnwidth]{Graphics/Concept_Fig_3.pdf}
	\includegraphics[width=1.9\columnwidth]{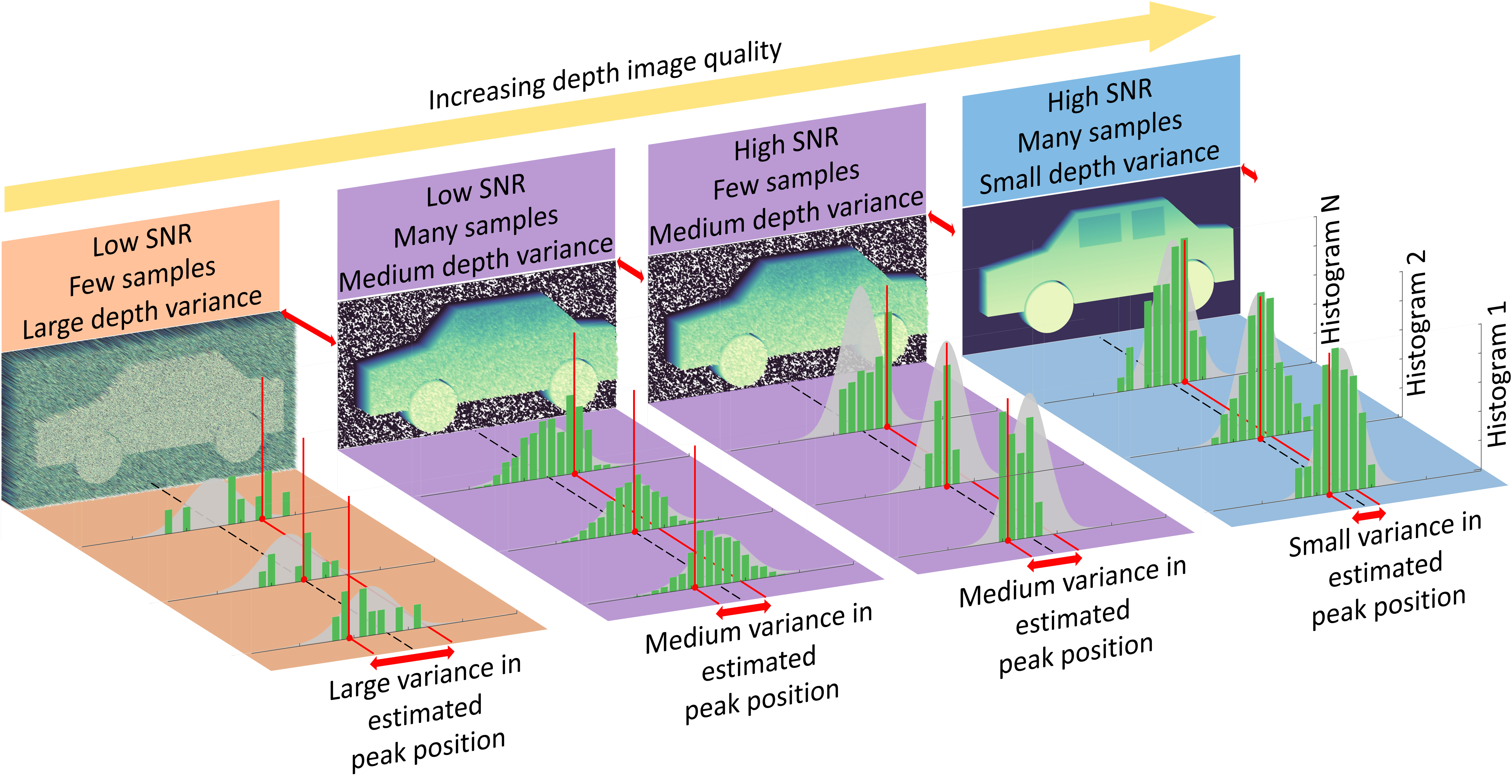}
	\caption{A conceptual illustration of the relationship between the variance in the estimate of the peak position of the histograms and the obtained depth image. Two Signal to Background-Noise Ratio (SbNR) and two sampling cases are depicted for a total of four regimes. Within each regime the underlying probability distribution of photon arrivals is shown by the grey curve. This curve is consistently centered on the true depth value shown by the black dashed line. The grey curve is sampled to produce a series of histograms, as shown by the green bars. For each histogram, the vertical red line illustrates the estimated position of the peak i.e., the depth. The variance in the estimate of the peak position, around the true depth value, is depicted by the double-headed red arrows. When this variance is large due to low SbNR and sparse sampling, as shown in the left most regime, the distinguishability in the depth image is poor i.e., fine features cannot be seen. When the variance is reduced, either by denser sampling or, higher SbNR (middle two regimes), the quality of the depth image improves. Lastly, when the histogram position can be repeatedly accurately determined (high SbNR with dense sampling) the variance is small and fine features in the depth image can be seen.}
        %\caption{A conceptual illustration of the relationship between the variance $\sigma$ in the position of the measured histograms (labelled here as peak position variance) and the obtained depth image (depicted as a vehicle). When the variance in the position of the histograms, (either between images for a single pixel, or, between adjacent pixels in the same image) is large, as shown by $\sigma_{1}$, the distinguishability in the depth image is poor. By contrast, when the histogram position can be repeatedly accurately determined i.e., a low variance $\sigma_{2}$, fine features in the depth image can be seen.}
	    \label{fig: regime_Fig}
\end{figure*}
Assuming that the impulse-response-function of the laser pulse and SPAD is approximately Gaussian in time with a mean $\mu$ and a standard deviation $\sigma'$, Eq.~\ref{eqn: photons per pulse} can be extended to a likelihood function $L(t|\mu,\sigma')$,
\begin{equation}
\begin{aligned}
    L(t|\mu,\sigma') = C_{dc} + C_{bckg} + \frac{P_{pp}}{\sigma'\sqrt{2\pi}}\exp{\left[-\frac{1}{2}\left(\frac{t-\mu}{\sigma'}\right)^2\right]}.
\end{aligned}
\label{eqn: Likelyhood}
\end{equation}
Here, $C_{dc}$ is the dark count rate of the detector measured in Hz, and $C_{bckg}$ is the background counts in Hz, both of which are constant in time. In the case of solar background, a solar photon which strikes the target must travel through the same photon channel as the signal photon in order to reach the detector, hence $C_{bckg}$ is
\begin{equation}
\begin{aligned}
    C_{bckg} = \frac{\lambda}{h c}\frac{q\Gamma e^{\frac{-R}{C_{atm}}}}{8 f_{no}^2}W_{bckg}W_pH_p,
\end{aligned}
\label{eqn: background per pulse}
\end{equation}
where $W_{bckg}$ is the solar background at $\lambda$ in Watts-per-square-meter. The standard deviation of the impulse-response-function is defined as $\sigma' = (\text{FWHM}/2\sqrt{2\ln{2}})$, where FWHM is the full-width at half-maximum of the impulse-response-function. Equation~\ref{eqn: Likelyhood} is defined such that
\begin{equation}
\begin{aligned}
    \int_{0}^{T} L(t|\mu,\sigma') dt = \alpha = T(C_{dc}+C_{bckg})+P_{pp},
\end{aligned}
\label{eqn: total photons}
\end{equation}
i.e., integrating the likelyhood for the Time Correlated Single Photon Counting (TCSPC) interval $[0,T]$ returns the average number of counts $\alpha$ measured by the detector in that interval. The Fisher information per pulse $F(t|\mu,\sigma')$ with respect to the peak position $\mu$ is defined as,
\begin{equation}
\begin{aligned}
    &F(t|\mu,\sigma') = \int_{0}^{T}\left\{\frac{\partial \ln{\left[\frac{L(t|\mu,\sigma')}{\alpha}\right]}}{\partial \mu}\right\}^2\frac{L(t|\mu,\sigma')}{\alpha} dt\\
    &= \int_{0}^{T}\frac{P_{pp}^{2}(t-\mu)^2\exp{\left[-\left(\frac{t-\mu}{\sigma'}\right)^2\right]}}{\sigma'^62\pi\alpha\left\{C_{dc} + C_{bckg} + \frac{P_{pp}}{\sigma'\sqrt{2\pi}}\exp{\left[-\frac{1}{2}\left(\frac{t-\mu}{\sigma'}\right)^2\right]}\right\}} dt\\
\end{aligned}
\label{eqn: fisher definition}
\end{equation}
for which a closed form analytical solution does not exist. By accumulating multiple frames, wherein each frame consists of many laser pulses (but at most one photon measurement) a histogram of photon arrival times can be created. A lower bound on the standard deviation $\sigma_{\mu}^{*}$ in the estimate of the peak position of this histogram~\cite{cramer1946mathematical,knee2014amplification}, and by extension the quality of the depth image (Fig.~\ref{fig: Concept_Fig}) is given by the Cram\'er-Rao bound
\begin{equation}
\begin{aligned}
    \sigma_{\mu}^{*} = \frac{1}{\sqrt{N[1-(1-\alpha)^{\eta\nu}] F(t|\mu,\sigma')}}.
\end{aligned}
\label{eqn: theory cr bound 2}
\end{equation}
Here, $N$ is the number of frames accumulated to create the histogram and $1-(1-\alpha)^{\eta\nu}$ is the probability of measuring at least one photon per frame for a frame of exposure time $\eta$ and a laser repetition rate $\nu$. Together, $N[1-(1-\alpha)^{\eta\nu}]$ represents the number of successful events in the histogram. Equation~\ref{eqn: theory cr bound 2} characterises the minimum possible standard deviation associated with estimating the depth of a single point. We can additionally apply a stricter criteria and define the minimum distinguishability $\sigma_{\mu}$ as $\sigma_{\mu} = \sigma_{\mu}^{*}\times2\sqrt{2\ln{(2)}}$ i.e., in a manner analogous to the Rayleigh resolution criteria, two depths are considered distinguishable when the peaks of the distributions associated with each point are separated by at least one FWHM.\\
\\
By examining Eqs.~\ref{eqn: photons per pulse}-\ref{eqn: theory cr bound 2} it is possible to isolate the effects of individual optical parameters on the variance in the depth estimation. Specifically, changes to the parameters that are shared between Eqs.~\ref{eqn: photons per pulse} and~\ref{eqn: background per pulse} do not effect the Signal to Background-Noise Ratio (SbNR) of the system and so have a negligible impact (if $C_{dc} << C_{bckg} + P_{pp}$) on the Fisher information per pulse (Eq.~\ref{eqn: fisher definition}). However, changing these parameters does effect the average number of counts per pulse $\alpha$ (Eq.~\ref{eqn: total photons}) and therefore the number of successful measurements $[1-(1-\alpha)^{\eta\nu}]$ per frame. Consequently, for fixed image acquisition time (finite $N$ and $\eta$) the imaging ability of the Lidar system is affected. Alternatively, the Fisher information can be acted upon directly by adjusting the SbNR given by SbNR $=[E_{0}\exp(-R/C_{atm})]/[W_{bckg}\pi R^2 \tan^2(\theta)]$. Further, from the integrand of Eq.~\ref{eqn: fisher definition}: when $P_{pp}$ = 0, the Fisher information is 0, which is consistent with the notion that depth cannot be inferred from dark counts and background photons arriving at the SPAD at random; and, the Fisher information is strongly governed by the pulse width $\sigma'$. This strong $\sigma'$ dependence in conjunction with the $E_0$ dependence in the SbNR implies that for the same average power, laser illuminators with higher peak powers will have superior imaging performance. Finally, the performance of the system can be affected by changes to the laser repetition rate $\nu$ in a manner comparable to changes in the $\alpha$ parameter. This is consistent with the intuition that the imaging performance of the Lidar system is contingent on the average power $\nu E_0$ of the laser illuminator.\\
\\
Figure~\ref{fig: regime_Fig} conceptually illustrates the imaging performance in the 4 regimes which arise as a result of the independence between the number of samples and the SbNR.  Specifically, two SbNR and two sampling cases are depicted. For low SbNR and sparse sampling the variance in the estimate of the peak position (Fig.~\ref{fig: regime_Fig} double-headed red arrows) is large and the resultant distinguishability in the depth image poor i.e., fine features cannot be seen. When the variance is reduced, either by denser sampling or higher SbNR, the quality of the depth image improves. Lastly, when the histogram position can be repeatedly accurately determined the variance is small and fine features in the depth image can be seen.\\
\\
The notion of distinguishability of the histogram peak is subtly, but importantly different from the notion of `resolution' (in the common definition). This is because a `high resolution sensor' with both a large number of transverse pixels and fine temporal sampling can still produce a poor quality depth image if used in a scenario where the peak of the histogram cannot be distinguished. A more apt comparison is `focus'. In an analogous manner to distinguishing two points at the same object plane in a traditional image (being `in focus'), a high quality depth image is one in which two points at different object planes (either between adjacent pixels, or, for the same pixel in consecutive images) can be reliably distinguished. For this reason we adopt the term `distinguishability' (rather than resolution) to describe the depth imaging abilities of a system for the remainder of this work.
\subsection{Simulating SPAD data}
\label{subsec: Simulating SPAD data}
The first regime in which the model can be employed is one where the Cram\'er-Rao bound is used directly to generate realistic depth images. The minimum distinguishability $\sigma_\mu$ derived from Eq.~\ref{eqn: theory cr bound 2} characterises the minimum possible standard deviation associated with estimating the peak of a histogram. Consequently, a depth image can be generated by taking a ground truth image i.e., an image of true depth $\mu$ values and noising each pixel $m,q$ according to $\mu_{m,q} \rightarrow \mu_{m,q}+\Delta_\mu$ where $\Delta_\mu$ represents a sample drawn from a normal distribution with a mean of $0$ and a standard deviation of the minimum distinguishability $\sigma_\mu$. As no histograms are produced in this method, it is highly computationally efficient and is able to generate $100 000$ images (with a transverse resolution of $192\times128$) in under 20 seconds, making it well suited to the generation of training datasets for machine learning applications. However, this method is subject to the assumption that the peak in the histograms can be estimated by a procedure that is capable of saturating the Cram\'er-Rao bound. Specifically, the Cram\'er-Rao bound, and by extension the minimum distinguishability $\sigma_\mu$, can only be reached by a minimum variance estimator and is only valid (as a lower bound) for unbiased estimators of the peak position~\cite{gyongy2020high}. Importantly, the existence of a minimum variance estimator (or even an unbiased estimator) is not guaranteed for all distributions of photon arrivals (Eq.~\ref{eqn: Likelyhood}). Consequently, the minimum distinguishability $\sigma_\mu$ represents the absolute best case performance of an imaging system independent of whether such performance can be achieved. The minimum distinguishability $\sigma_\mu$ remains beneficial as it represents the absolute best case scenario for any instance of an imaging system and acquisition time.\\
\\
The second regime in which the model can be employed is one where realistic histograms are generated via a process that simulates the photon-by-photon detection of a SPAD array. This approach addresses any dependence on minimum variance estimators as depths can now be measured using any appropriate peak estimation method applied directly to synthetic data. This `histogram' approach provides a highly flexible means of simulating the performance of the imaging system for situations in which there is a large difference between the distinguishability which can actually be obtained and the Cram\'er-Rao bound. i.e., situations for which unbiased estimators of Eq.~\ref{eqn: Likelyhood} do not exist. The cost of this approach is computational complexity with a single image requiring $\sim1$ hour of computation. It should be noted though that this computation time is dependent upon the number of pixels, the number of frames-per-image, and the number of laser-pulses-per-frame.\\ 
\\
Consider a single pixel of a SPAD detector having $b$ bins of width $\omega$ for a total TCSPC interval of $T = b\omega$. The probability vector $P_{i}^{'}$ for the arrival of a photon in a bin for a given laser pulse is,
\begin{equation}
\begin{aligned}
    P_{i}^{'} &= \int_{[(i-1)\omega,i\omega)}^{}L(t|\mu+j,\sigma')dt|_{i\in\mathbb{Z}_{[1,b]}^{+};j\sim\mathcal{N}(\mu_{j},\sigma_j)},\\
    &=\omega(C_{dc}+C_{bckg})+\frac{P_{pp}}{2}\Biggl\{\text{erf}\left[\frac{\mu+j-(i-1)\omega}{\sigma'\sqrt{2}}\right]-\\
    &\text{erf}\left(\frac{\mu+j-i\omega}{\sigma'\sqrt{2}}\right)\Biggr\}|_{i\in\mathbb{Z}_{[1,b]}^{+};j\sim\mathcal{N}(\mu_{j},\sigma_j)},
\end{aligned}
\label{eqn: histo prob}
\end{equation}
i.e., the probability of a photon in the i\textsuperscript{th} bin is the area under the likelyhood (Eq.~\ref{eqn: Likelyhood}) inclusive of the bins' lower bound ($(i-1)\omega$) and exclusive of its upper bound ($i\omega$). $j$ represents the jitter associated with a given pulse as a displacement of $\mu$ sampled from a Normal distribution $\mathcal{N}$ having a mean $\mu_j > 0$ with a standard deviation $\sigma_j$.\\
\begin{figure}[t!]
\centering
    %\begin{adjustbox}{max width=\columnwidth,center}
	\includegraphics[width=0.95\columnwidth]{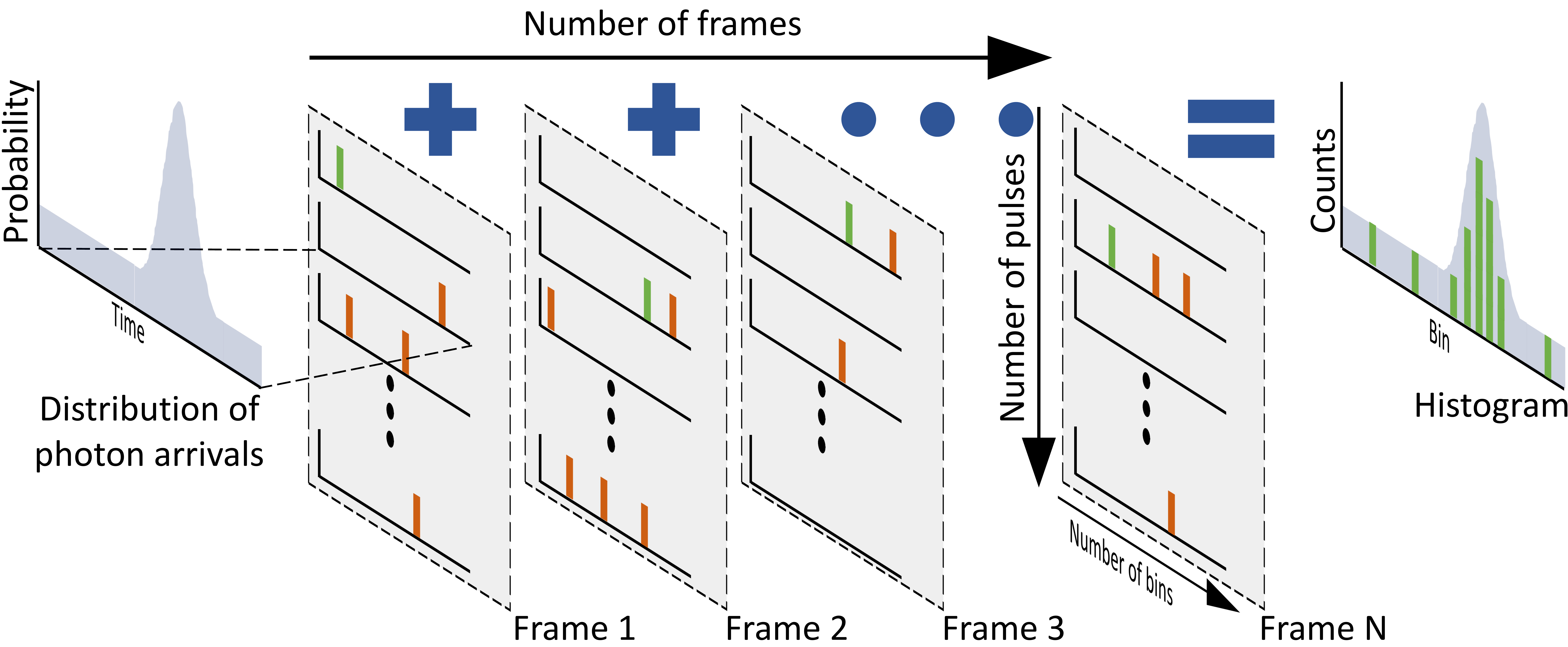}
        \caption{A conceptual illustration of the SPAD histogram simulation process. The salient parameters of the SPAD imaging system (Fig.~\ref{fig: Concept_Fig}) are used to define a probabilistic distribution of photon arrivals. Next, a data structure with a dimensionality corresponding to the number of bins $b$, the number of laser pulses $\eta\nu$, and the desired number of frames $N$ is populated by a Binomial sampling procedure based on Eq.~\ref{eqn: histo prob}. The index of the first non-zero value (green bars) on a per-frame basis is then recorded with subsequent non-zero values (orange bars) ignored precisely simulating the `first-photon behaviour' of the SPAD. The first non-zero values can then be accumulated over multiple frames to reconstruct a histogram with a distribution matching that of  Eq.~\ref{eqn: histo prob}.}
	    \label{fig: SPAD_sim}
\end{figure}
\\
To simulate SPAD histograms the parameters of the optical system (bulleted points in Fig.~\ref{fig: Concept_Fig}) are used to define a probabilistic distribution of photon arrivals on a per-pulse basis. A data structure having a dimensionality corresponding to the number of bins $b$, the number of laser pulses (as determined by the exposure time $\eta$ and the laser repetition rate $\nu$), and the desired number of frames $N$ is then created. This data structure is populated by a Binomial sampling process that treats each element (i.e., each histogram bin) as an independent single-trial event with a probability determined by Eq.~\ref{eqn: histo prob}. In this way we fully model (for a single pixel) all possible photon arrivals on a per-bin per-pulse basis as depicted conceptually in Fig.~\ref{fig: SPAD_sim}. This process is highly parallaziable and has been optimized to run on a GPU. Once populated, the data structure is processed such that only the bin index of the first non-zero value (i.e., the first photon) on a per-frame basis is recorded precisely simulating the `first-photon behaviour' of the SPAD. These indices can then be accumulated over the number of frames to construct a histogram. Further, since the Binomial distribution converges with the Poisson distribution in the many trial limit~\cite{novak2011extreme}, our approach remains consistent with prior works whilst extending the simulation to the single-bin, single-trial limit.\\
\\
To extend Eq.~\ref{eqn: histo prob} to an entire image (from a single pixel) we model $\mathcal{M}\times\mathcal{Q}$ independent (neglecting cross talk) pixels and introduce an additional noise parameter $k$ which captures the inter-pixel variance at an inter-frame scale. $k$ is a sensor specific parameter which characterises the random discrepancies in the triggering of timing circuitry between pixels as a result of manufacturing variations in the propagation of the timing trigger signal across the sensor~\cite{henderson2019192}. The intra-pixel noise modifies Eq.~\ref{eqn: histo prob} such that the probability associated with pixel $m,q$ is
\begin{equation}
\begin{aligned}
    P_{i,m,q}^{'} &= \int_{[(i-1)\omega,i\omega)}^{}L(t|\mu+j+k,\sigma')dt|_{i\in \mathbb{Z}_{[1,b]}^{+};j\sim\mathcal{N}(\mu_{j},\sigma_j);k\sim\mathcal{N}(0,\sigma_q)}.
\end{aligned}
\label{eqn: histo prob img}
\end{equation}
\section{Experimental results and discussion}
\label{sec: Experimental results}
\subsection{Resolution test target}
\label{subsec: Resolution test target}
\begin{figure}[t!]
    %\begin{adjustbox}{max width=\columnwidth,center}
	\centering
	    \setlength\tabcolsep{3pt}
	    \footnotesize
        \begin{tabular}{c c}
            \toprule
            \textbf{Reference image}&\textbf{Ideal depth}\\
            \cmidrule(lr){1-1}
            \cmidrule(lr){2-2}
            \includegraphics[width=0.45\columnwidth]{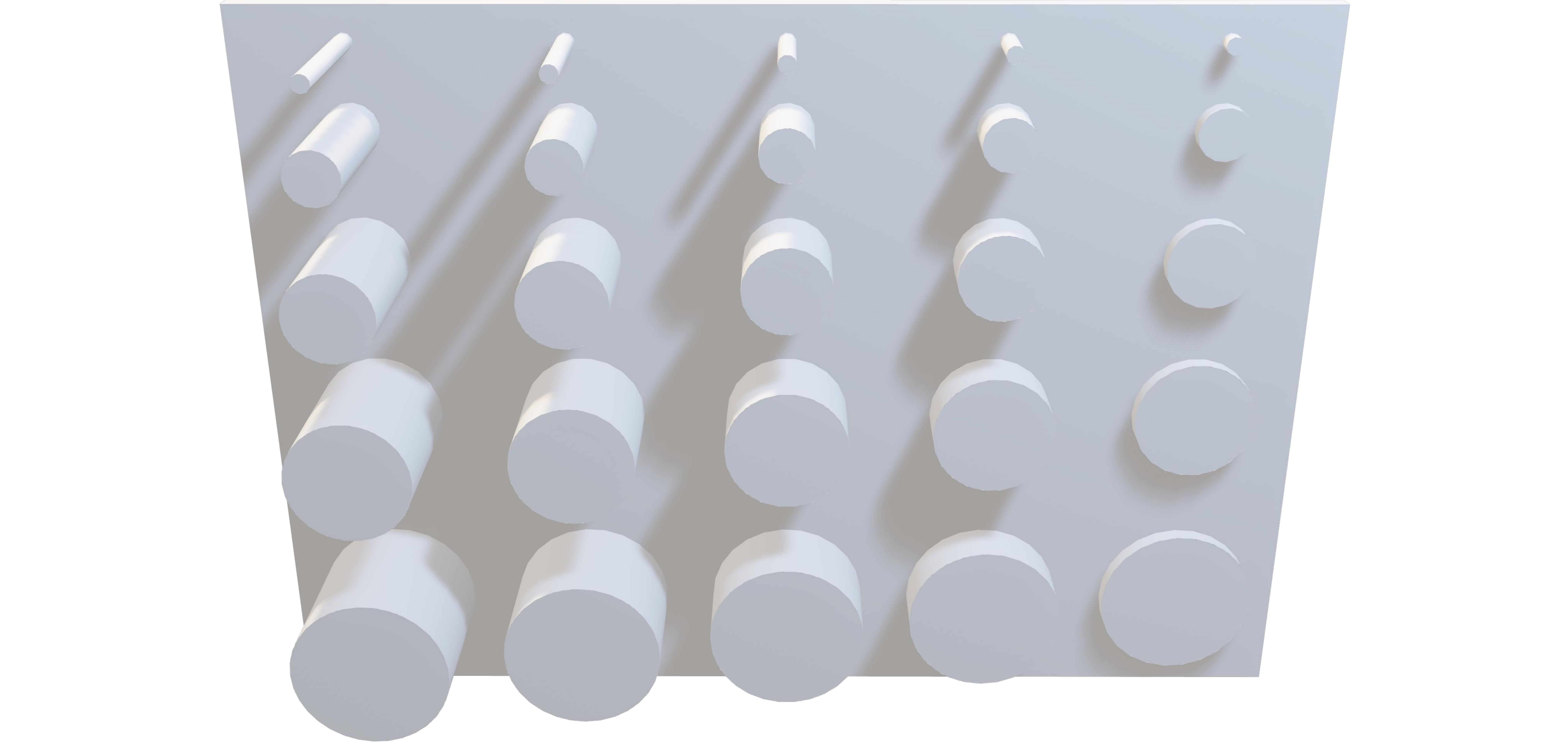}&\includegraphics[width=0.45\columnwidth]{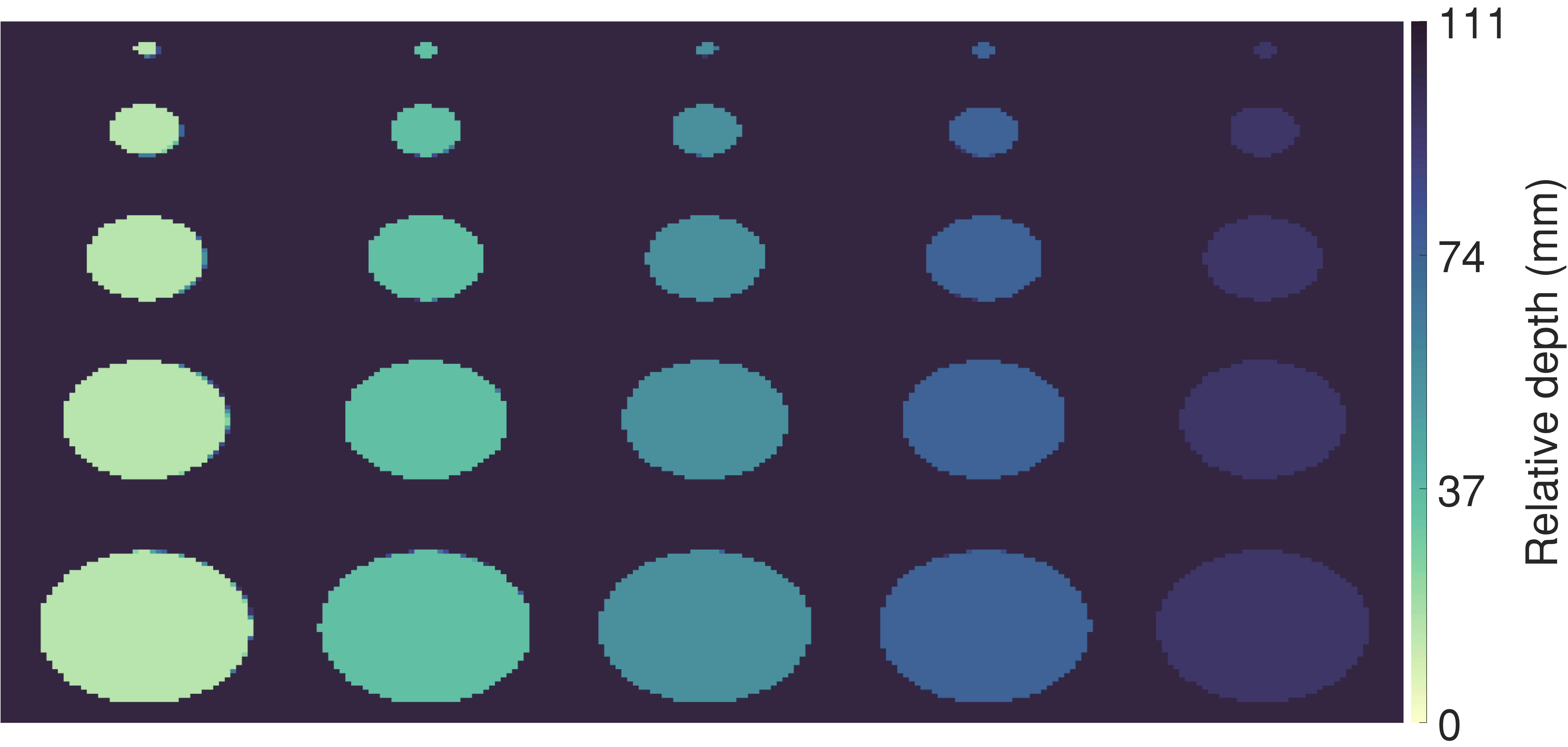}\\
        \bottomrule
        \end{tabular}
        \captionof{figure}{A reference image of the resolution test target. The test target consisted of a series of cylinders ranging in both diameter and height from 90 mm to 10 mm in 20 mm increments. The ground truth depth map for the simulation of the resolution test target. }
	    \label{fig: Res_Target_base}
\end{figure}
To validate the predictions of Sec.~\ref{sec: Theory} the resolution test target shown in Fig.~\ref{fig: Res_Target_base} was constructed. The test target consisted of a white backboard to which equidistantly spaced white plastic cylinders ranging in both diameter and height from 90 mm to 10 mm in 20 mm increments were attached. The target was placed at a range of $R\sim15$ m and illuminated using a laser operating at $\lambda = 671$ nm and $2.25$ MHz with a pulse energy of $E_0 = 1$ nJ. For the full list of experimental parameters the reader is referred to Sec.~\ref{subsec:Experimental parameters used in simulating the resolution test target}. The target was imaged using a state-of-the-art SPAD consisting of 192$\times$128 pixels in a 2:1 aspect ratio with 4096 50-picosecond bins per-pixel and a one-photon-per-frame sensing limit~\cite{henderson2018192}. 
\begin{figure}[b!]
    %\begin{adjustbox}{max width=\columnwidth,center}
	\centering
	    \setlength\tabcolsep{3pt}
	    \footnotesize
        \begin{tabular}{c c c c}
            \toprule
            %&Reference image&Ideal depth\\
            %\cmidrule(lr){2-2}
            %\cmidrule(lr){3-3}
            %&\includegraphics[width=0.4\columnwidth]{Graphics/Res_Target_Ideal_Ref.pdf}&\includegraphics[width=0.4\columnwidth]{Graphics/Res_Target_Ideal.pdf}&\\
            %\multicolumn{1}{c}{\bfseries OAM free annulus}\\
            %\cmidrule[\heavyrulewidth](lr){1-3}
            &\textbf{Distributions}&\multicolumn{2}{c}{\textbf{Depth}}\\
            &\textbf{of depths}&\multicolumn{2}{c}{\textbf{images}}\\
            \cmidrule(lr){2-2}
            \cmidrule(lr){3-4}
            \begin{turn}{90}\hspace{1.25cm}\textbf{CRB}\end{turn}&\includegraphics[width=0.3\columnwidth]{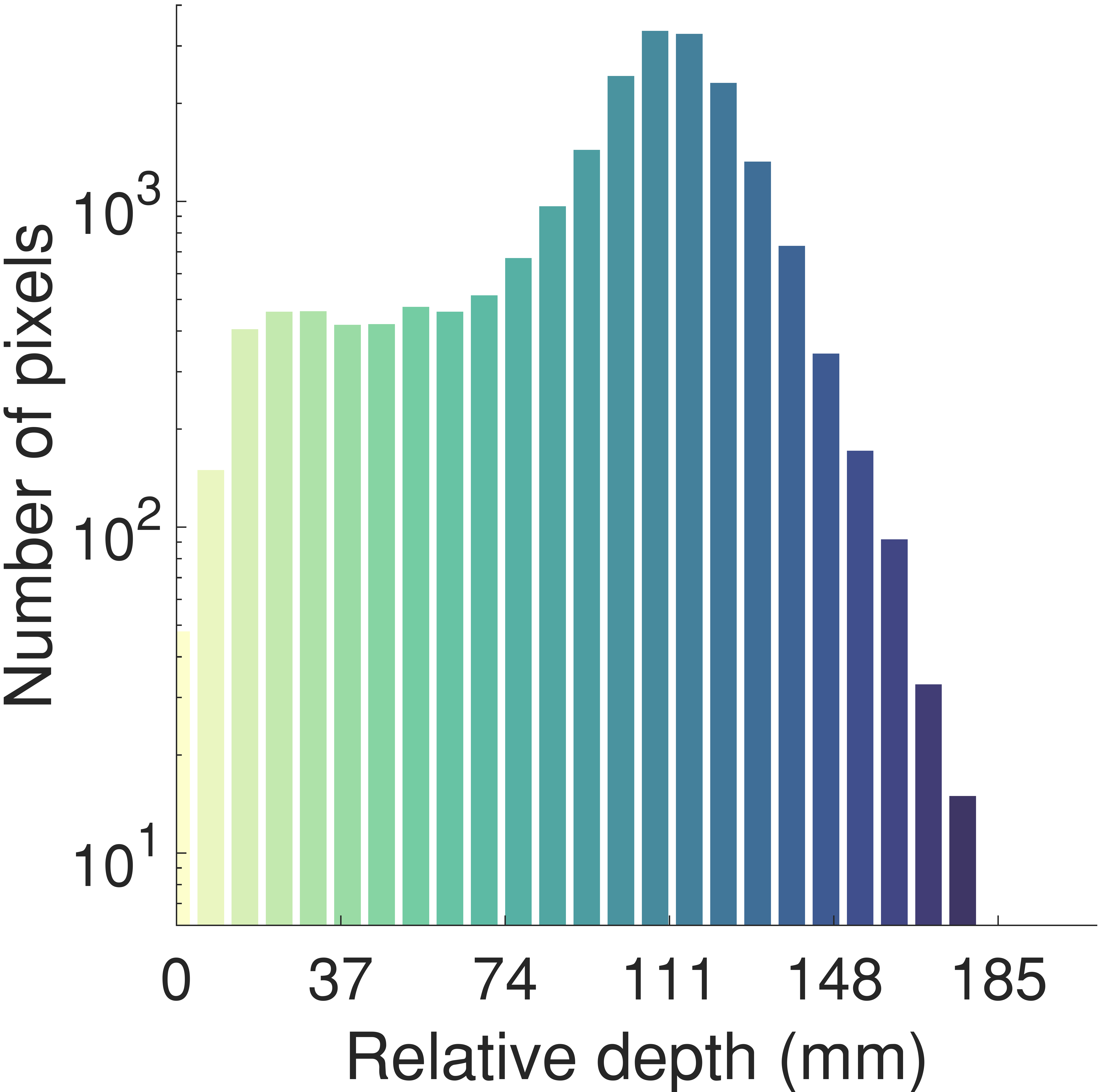}&\multicolumn{2}{c}{\includegraphics[width=0.6\columnwidth]{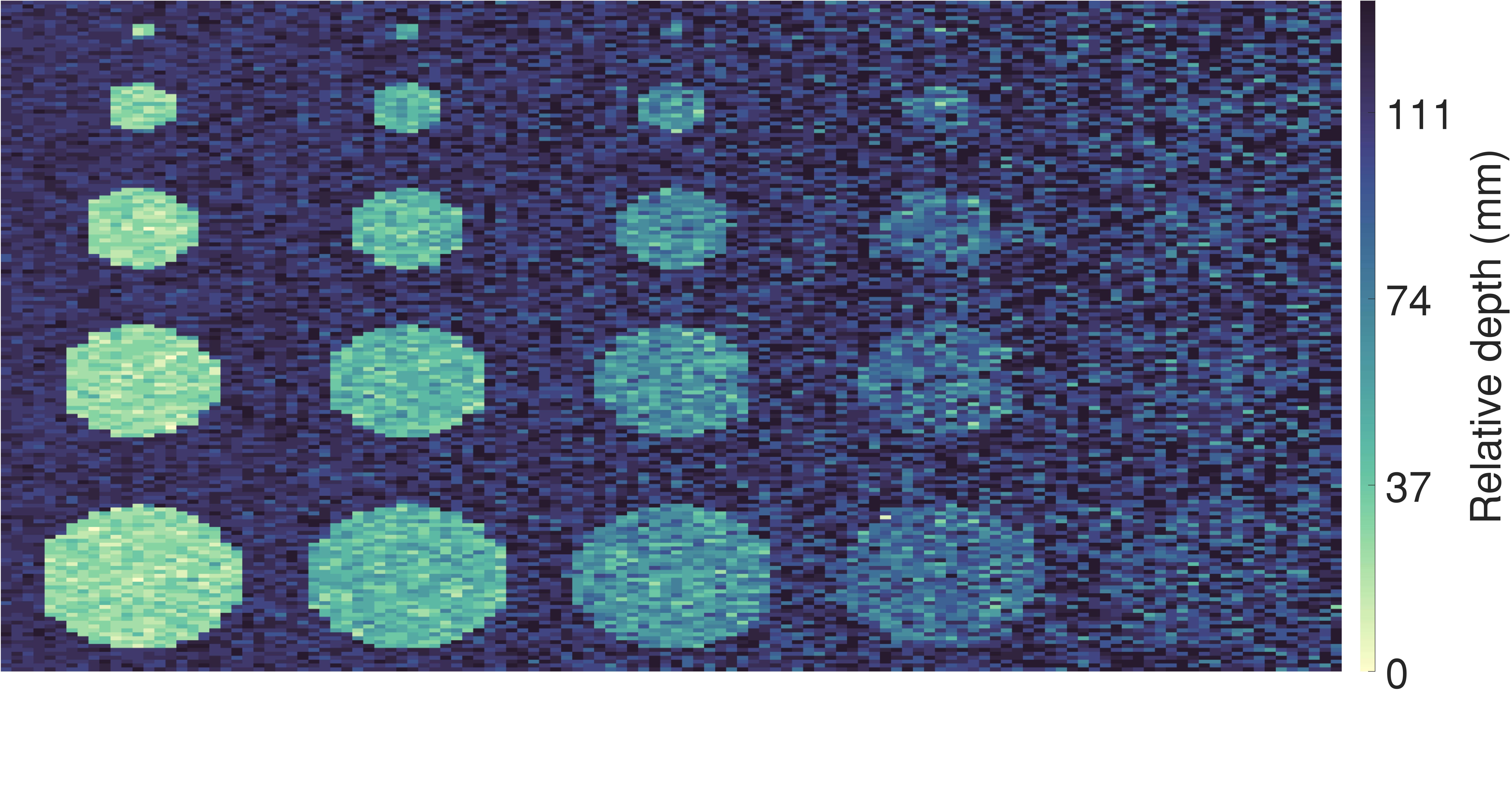}}\\
            \begin{turn}{90}\hspace{1cm}\textbf{Histograms}\end{turn}&\includegraphics[width=0.3\columnwidth]{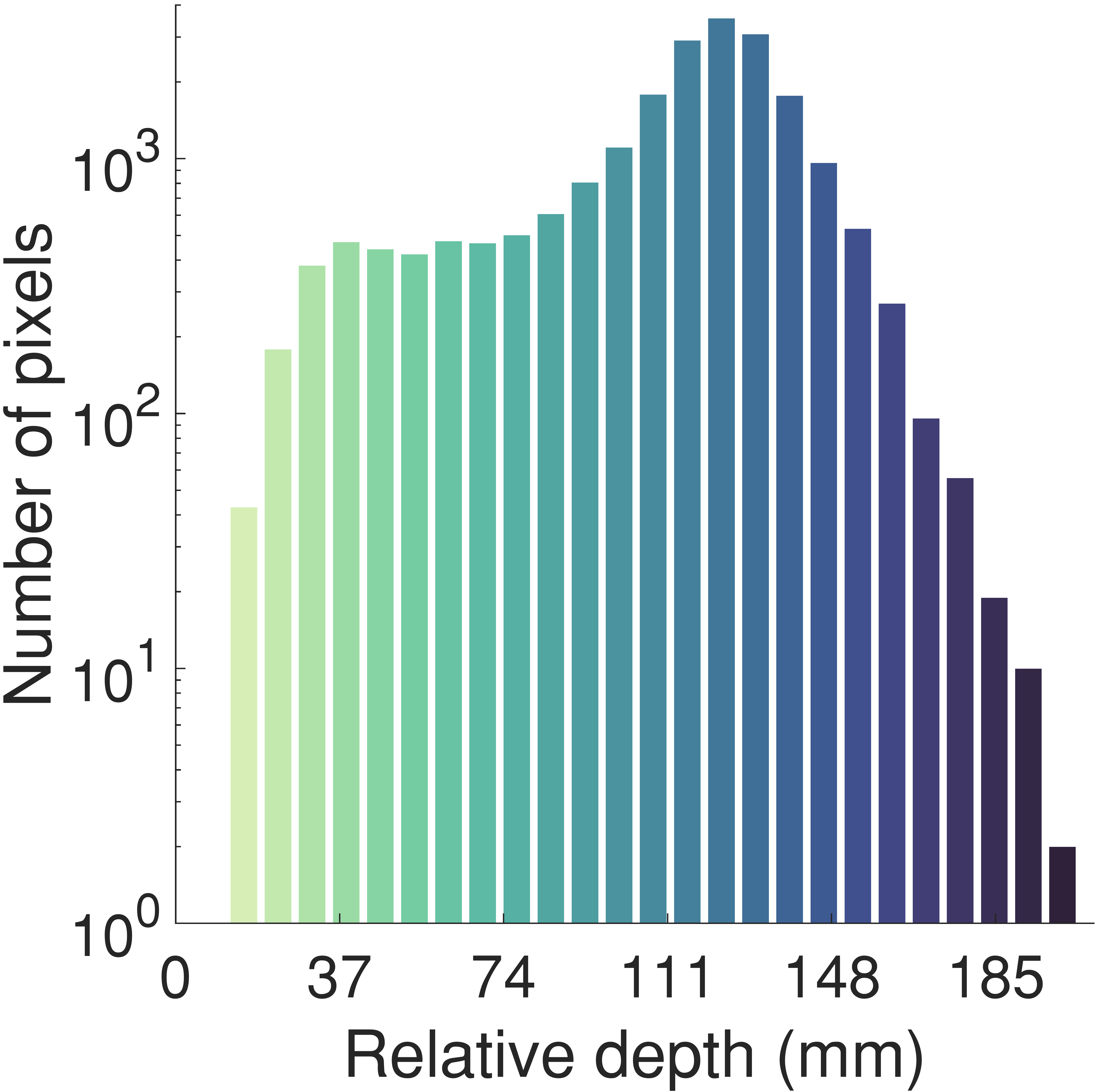}&\multicolumn{2}{c}{\includegraphics[width=0.6\columnwidth]{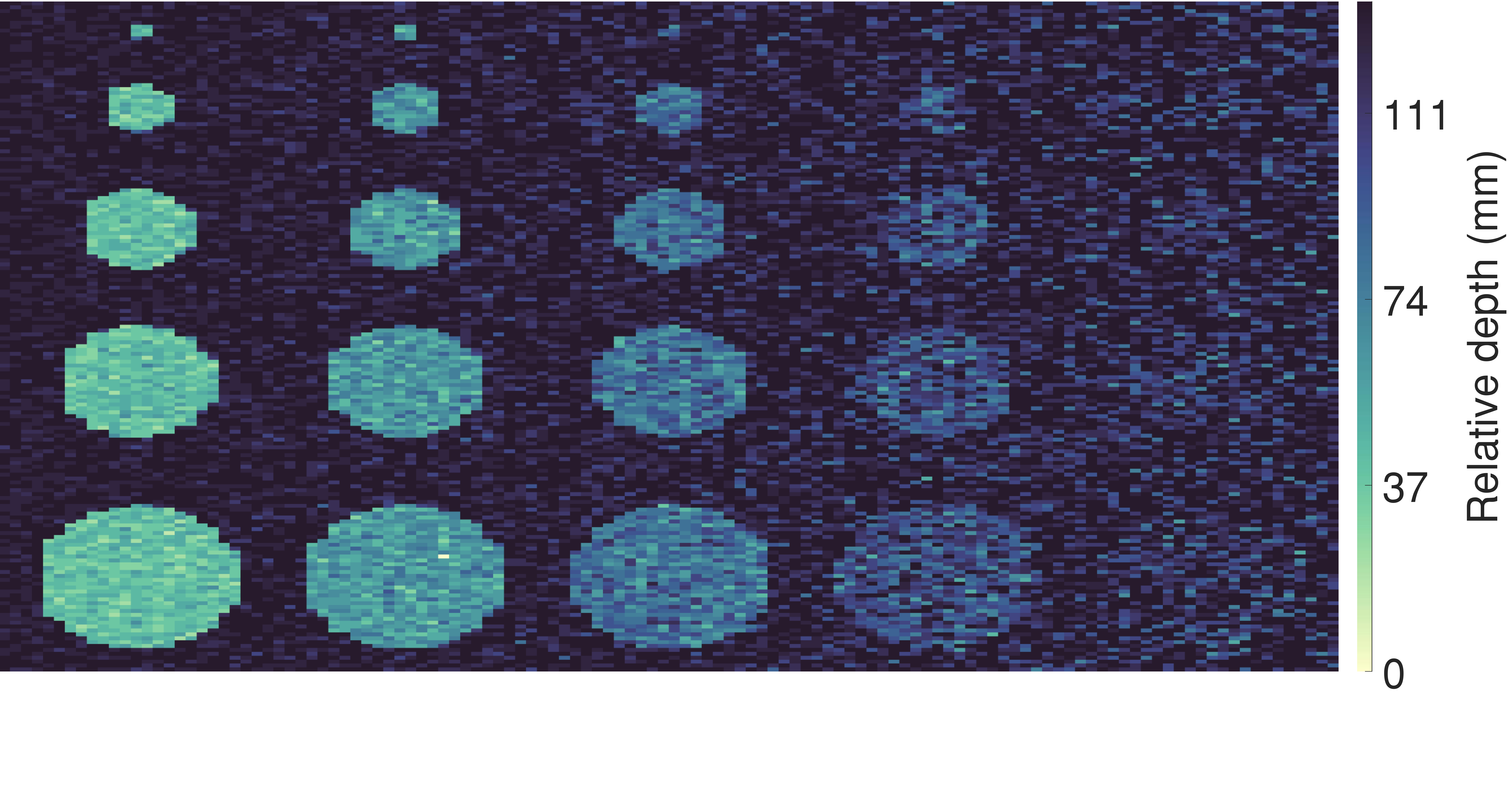}}\\
            \cmidrule(lr){2-4}
            \begin{turn}{90}\hspace{1cm}\textbf{Experiment}\end{turn}&\includegraphics[width=0.3\columnwidth]{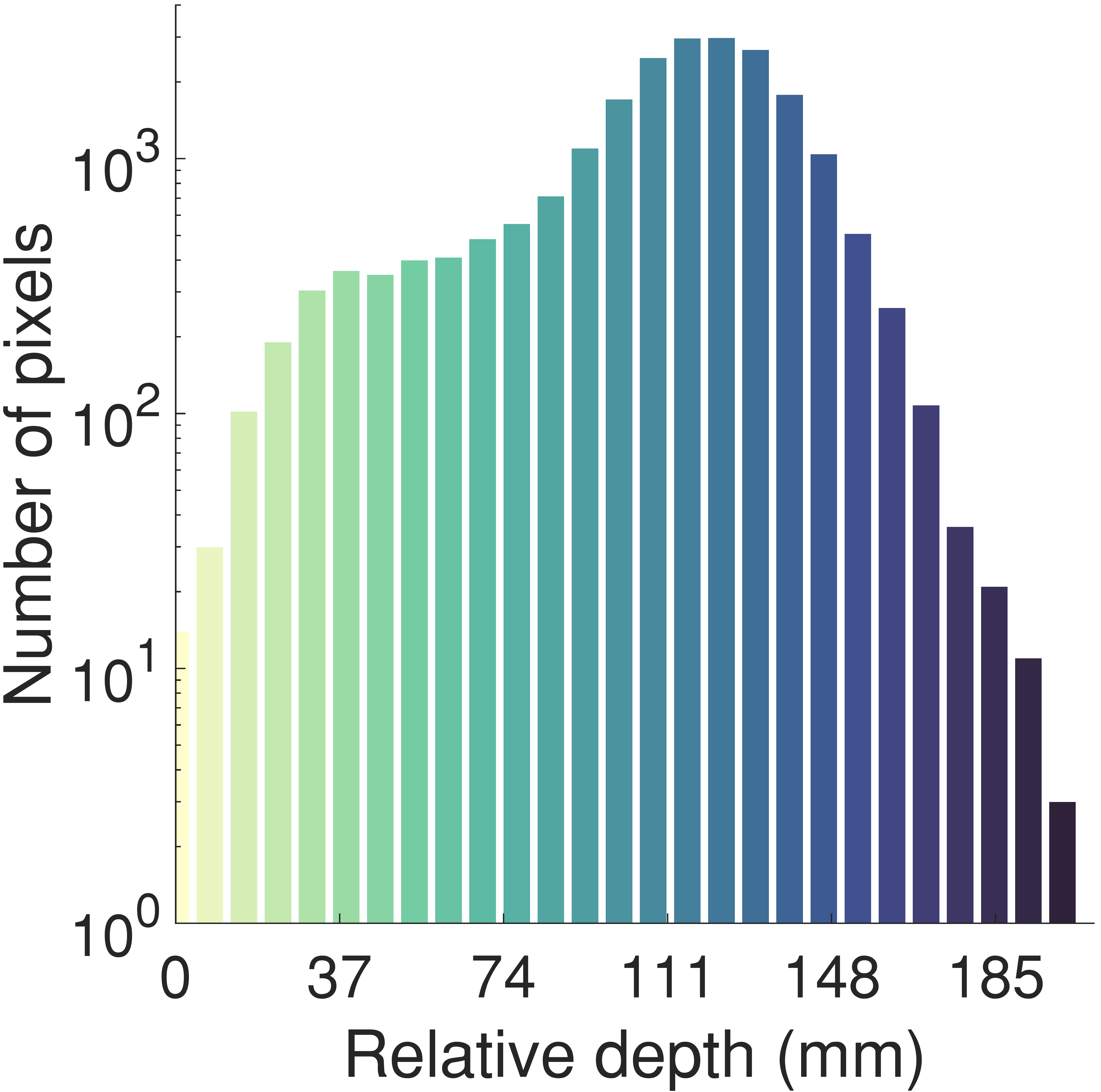}&\multicolumn{2}{c}{\includegraphics[width=0.6\columnwidth]{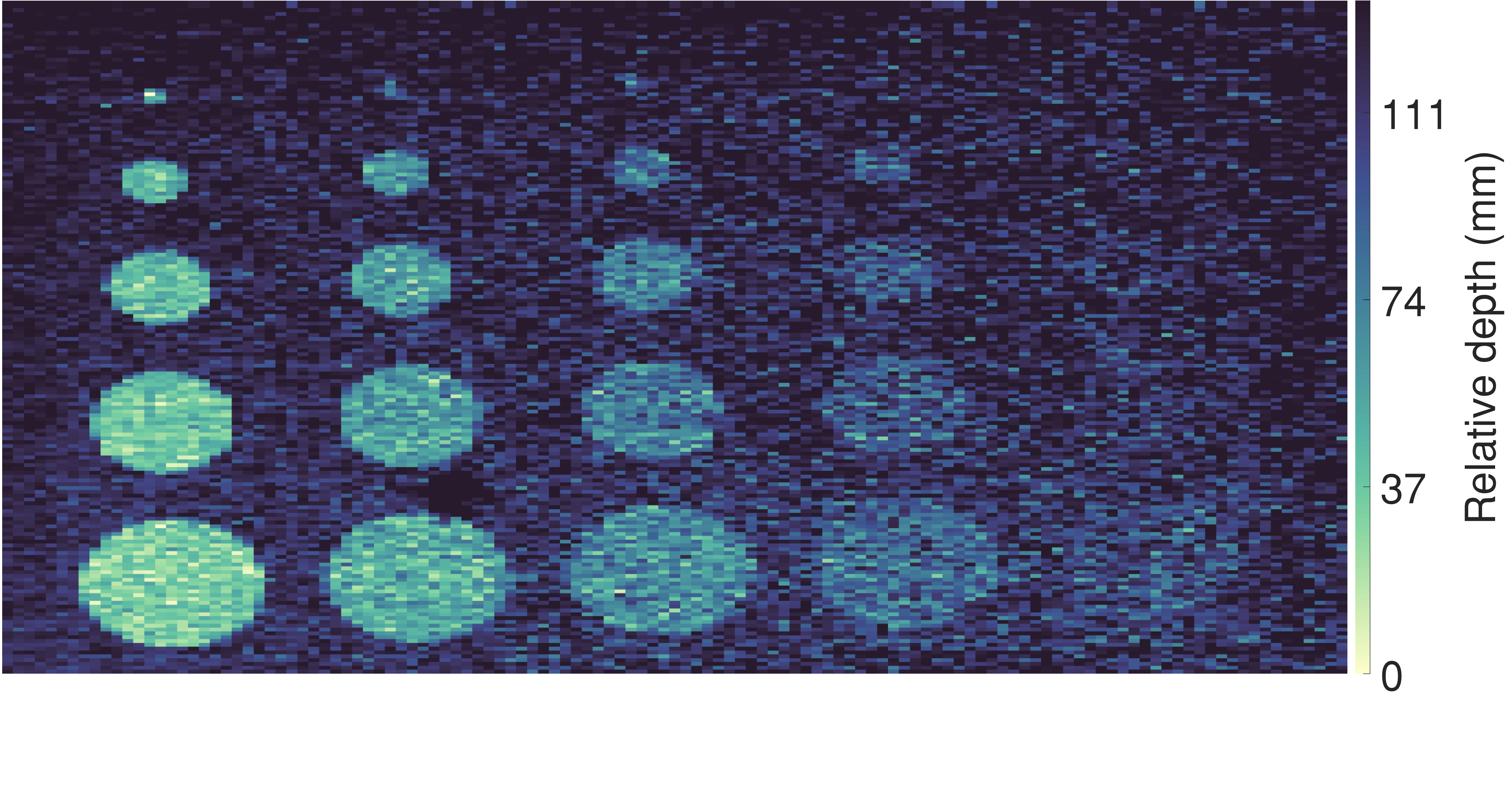}}\\
        \bottomrule
        \end{tabular}
        \captionof{figure}{Quantitative and qualitative comparison between the simulated and measured depth images for the resolution test target. Left column, the number of pixels at a given depth as a histogram. Right column, the depth images. The first and second rows show the simulated images calculated using the Cram\'er-Rao bound approach and the histogram simulation approach respectively. The bottom row shows the experimental results.}
	    \label{fig: Res_Target}
\end{figure}
A 50 mm focal length lens was used to image the target such that the field-of-view of the sensor matched the transverse size of the test target resulting in a transverse pixel resolution at the target plane of $4.55\times2.22$ millimeters-per-pixel. Consequently, the narrowest posts of the resolution test target ($10$ mm) are represented by only $\sim2\times4$ pixels, highlighting the ongoing need to develop high transverse resolution sensors. The inter-pixel noise parameter $k$ was determined to follow a Normal distribution with $\mu_{k} = 0$ and a standard deviation which varied linearly across the sensor such that for the first column of pixels $k\sim\mathcal{N}(0,\sigma_{q = 0})|\sigma_{q = 0} = 41\times10^{-12}$ whilst for the last column $k\sim\mathcal{N}(0,\sigma_{q = \mathcal{Q}})|\sigma_{q = \mathcal{Q}} = 166\times10^{-12}$.\\
\\
Figure~\ref{fig: Res_Target} shows the simulated and measured depth images of the resolution test target accumulated over 1000 frames using an f2 collection lens. Each frame had a 1 ms duration resulting in 2250 laser pulses-per-frame. Additionally, Fig.~\ref{fig: Res_Target} provides histograms showing the number of pixels at a given depth. The excellent qualitative agreement between the simulated and measured depth images demonstrates the ability of the model to accurately simulate SPAD data. The agreement is emphasised by the overlap between the depth histograms which have a median per bar accuracy in excess of $80\%$, meaning that under idealized lab conditions the simulation is able to predict the number of pixels at a given depth over an entire image with accuracies in excess of $80\%$.\\
\begin{figure}[b!]
\centering
    %\begin{adjustbox}{max width=\columnwidth,center}
	    \setlength\tabcolsep{2pt}
	    \footnotesize
        \begin{tabular}{c c}
	        \begin{turn}{90}\hspace{3.2cm}\textbf{f2 lens}\end{turn}&\includegraphics[width=0.95\columnwidth]{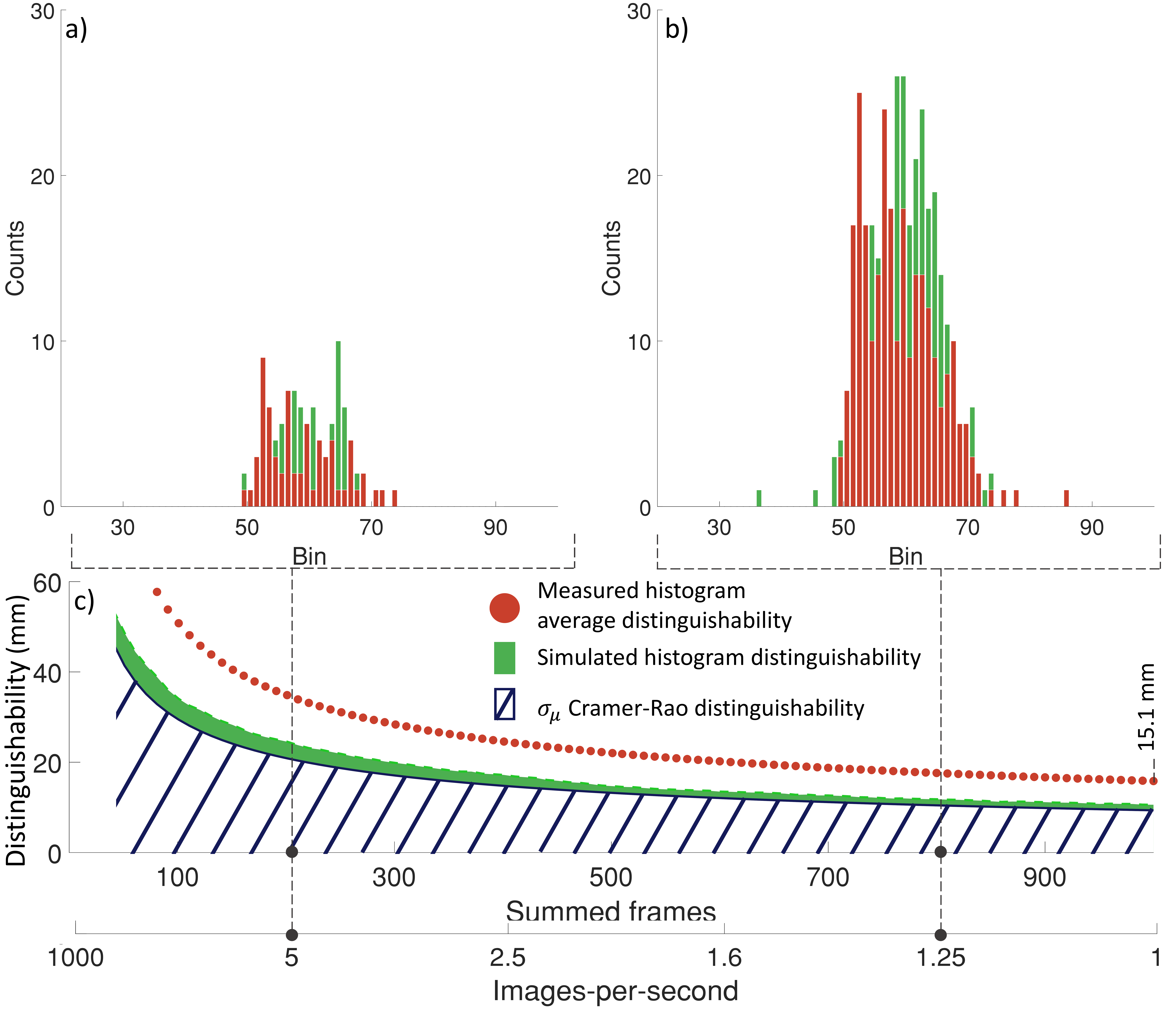}\\
	        \cmidrule(lr){1-2}
	        \begin{turn}{90}\hspace{1.25cm}\textbf{f4 lens}\end{turn}&\includegraphics[width=0.95\columnwidth]{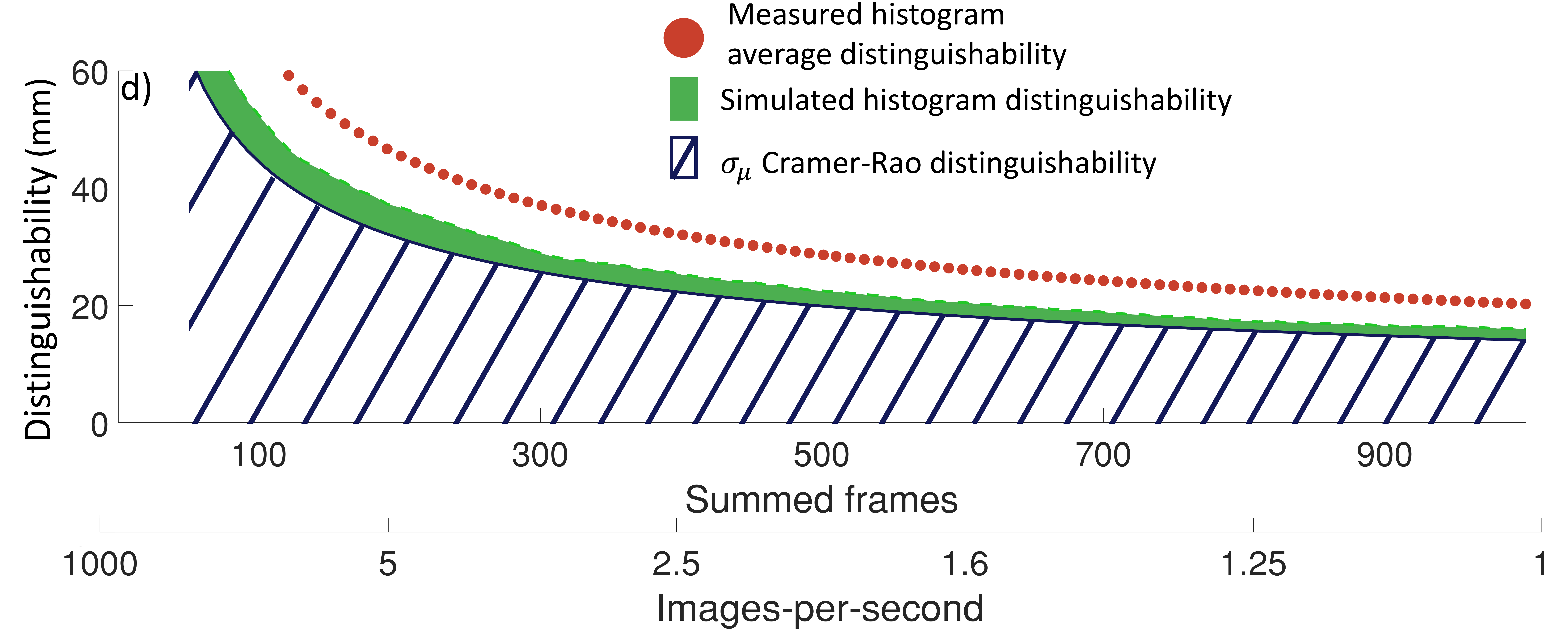}\\
        \bottomrule
        \end{tabular}
        \captionof{figure}{The simulated minimum distinguishability based on Eqs.~\ref{eqn: theory cr bound 2} and~\ref{eqn: histo prob} overlayed with the experimentally obtained average distinguishability of the resolution test target (Fig.~\ref{fig: Res_Target} bottom row).  c) the lower bounds on distinguishability as determined by Eqs.~\ref{eqn: theory cr bound 2} and~\ref{eqn: histo prob} shown in blue and green respectively for the case of an f2 lens. The average distinguishability based on experimental histogram variance is shown as orange dots. The upper sub-panels, a) and b), show exemplar simulated histograms (green) with experimentally obtained histograms (for a single pixel) overlayed in orange at the positions indicated by the risers. d) the lower bounds on distinguishability as determined by Eqs.~\ref{eqn: theory cr bound 2} and~\ref{eqn: histo prob} shown in blue and green respectively with the experimentally obtained average distinguishability overlayed as orange dots for the case of an f4 lens. At less than 50 summed frames the simulated histograms contain no counts and the minimum distinguishability becomes ill-defined.}
	    \label{fig: SPAD_Fisher_Exp}
\end{figure}\\
To create the images in Fig.~\ref{fig: Res_Target} the peak of the histogram associated with each pixel was determined using match filtering (a.k.a cross correlation). Specifically, a Gaussian kernel was convolved with the histograms and the position of highest response taken as the depth. We selected match filtering as an estimator due to its prevalence within Lidar data processing. Additionally, it represents an unbiased (although possibly not minimum variance) estimator for the case of a single Gaussian peak with a uniform background. For further discussion, the reader is directed to Sec.~\ref{subsec:Match filtering as an unbiased estimator}. Whilst more advanced processing techniques exist for depth calculation~\cite{kuzmenko20203d}, we stress that the objective of this work was not to examine the efficacy of different image processing techniques but was rather to accurately simulate the data acquired by SPAD based imaging systems.\\
\\
To further demonstrate the performance of the model we examined the distinguishability of a single pixel as a function of images-per-second, and lens f-number. Figure~\ref{fig: SPAD_Fisher_Exp} shows the simulated minimum distinguishabilities for a single pixel and directly compares these to the measured distinguishablilities of the resolution test target. The simulation is performed for a collection lens with f number $f_{no} = 2$ as well as $f_{no} = 4$. 1000 frames of SPAD data are simulated with an exposure time of $1$ ms (1000 fps). The number of frames $N$ accumulated to produce a histogram is varied from 1 (1000 images-per-second) to 1000 (1 image-per-second) in 100 increments and this process repeated 100 times to create 100 independent images at each frame summation increment.  The distinguishability of the simulated histograms is calculated as $2\sqrt{2\ln{(2)}}\times$ the standard deviation in the depth estimate of the 100 images at each summation increment. This process was then matched experimentally. The Fisher information is $F = 1.525\times10^{19}$ s$^{-2}$ and $F = 1.507\times10^{19}$ s$^{-2}$ for the f2 and f4 collection lenses respectively.\\
\\
Figure~\ref{fig: SPAD_Fisher_Exp} c) and d) show that as more frames are accumulated to produce the histogram, the distinguishability of the depth estimate (of both the Cram\'er-Rao bound and the histograms) improves at the expense of images-per-second. This improvement follows a proportionality of $1/N$ consistent with Eq.~\ref{eqn: theory cr bound 2}. However, the distinguishability of the histograms remains larger than that of the Cram\'er-Rao bound as a result of the match filtering used for peak estimation. The $1/N$ dependence confirms that the minimum axial distance that can be resolved, i.e., the distinguishability, of a SPAD based Lidar systems is contingent on both the imaging optics and the acquisition time. This is illustrated by comparing Fig.~\ref{fig: SPAD_Fisher_Exp} c) and d) which share the $1/N$ trend, but which have different absolute values due to the proportionally poorer light gathering capability ($[1-(1-\alpha)^{\eta\nu}]$) of the f4 collection lens. Further, Fig.~\ref{fig: SPAD_Fisher_Exp} c) and d) show that the distinguishability of the measured data improves with the same $1/N$ trend as predicted by Eq.~\ref{eqn: theory cr bound 2}. However, it remains poorer than the simulated histogram distinguishability in both the f2 and f4 lens cases despite the simulated and measured histograms ( Fig.~\ref{fig: SPAD_Fisher_Exp} a) and b)) close agreement. Additionally, the minimum measured distinguishability associated with the f2 lens (Fig.~\ref{fig: SPAD_Fisher_Exp} c)) is $15.1$ mm. This result is in keeping with the measured depth image in Fig.~\ref{fig: Res_Target} where the right-most column of $10$ mm cylindrical posts cannot be reliably distinguished from the back plane.\\ 
\\
The difference between the simulated and measured distinguishabilities in Fig.~\ref{fig: SPAD_Fisher_Exp} can be attributed to the $k$ parameter and its subsequent effects on Eq.~\ref{eqn: histo prob}. Specifically, Eq.~\ref{eqn: histo prob} (from which the single pixel simulated histograms are derived) characterises a single pixel from a perfectly uniform sensor, i.e., if Eq.~\ref{eqn: histo prob} was extended to an $\mathcal{M}\times\mathcal{Q}$ sensor it implicitly assumes perfect triggering between adjacent pixels. Consequently, the ability to distinguish the depths associated with two adjacent points at the target plane using a sensor described by Eq.~\ref{eqn: histo prob} would only be constrained by the completeness of the histograms. By contrast, the inclusion of the $k$ parameter in Eq.~\ref{eqn: histo prob img} allows for adjacent pixels to trigger at different times. Hence, even if the distinguisability of each individual pixel in a sensor described by Eq.~\ref{eqn: histo prob img} is small (i.e., the histograms are complete) the sensors ability to produce a depth image will be poor. This is because although the depth of adjacent points at the target plane could be calculated precisely, the accuracy would be poor, resulting in a distorted surface profile. This result implies that the quality of the surface profiles obtained by SPAD imaging systems is dependent not only on the temporal resolution of the individual pixels, but also on the consistency of that resolution across all pixels in the sensor. 
   
\subsection{Landrover}
\label{sec: Landrover}
\begin{figure}[b!]
    %\begin{adjustbox}{max width=\columnwidth,center}
	\centering
	    \footnotesize
        \begin{tabular}{c c}
            \toprule
            \textbf{Reflectivity map}&\textbf{Ideal depth}\\
            \cmidrule(lr){1-1}
            \cmidrule(lr){2-2}
            \includegraphics[width=0.45\columnwidth]{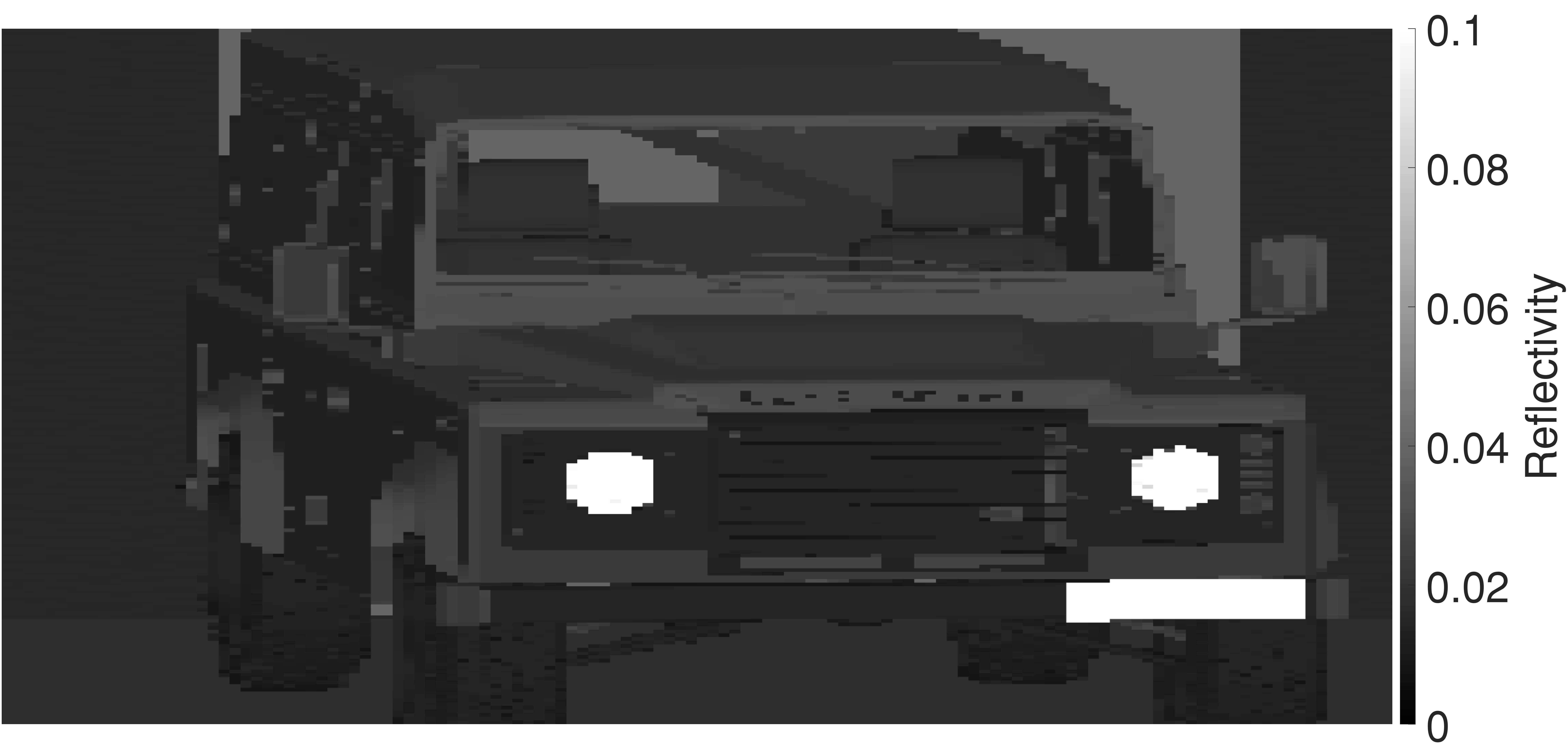}&\includegraphics[width=0.45\columnwidth]{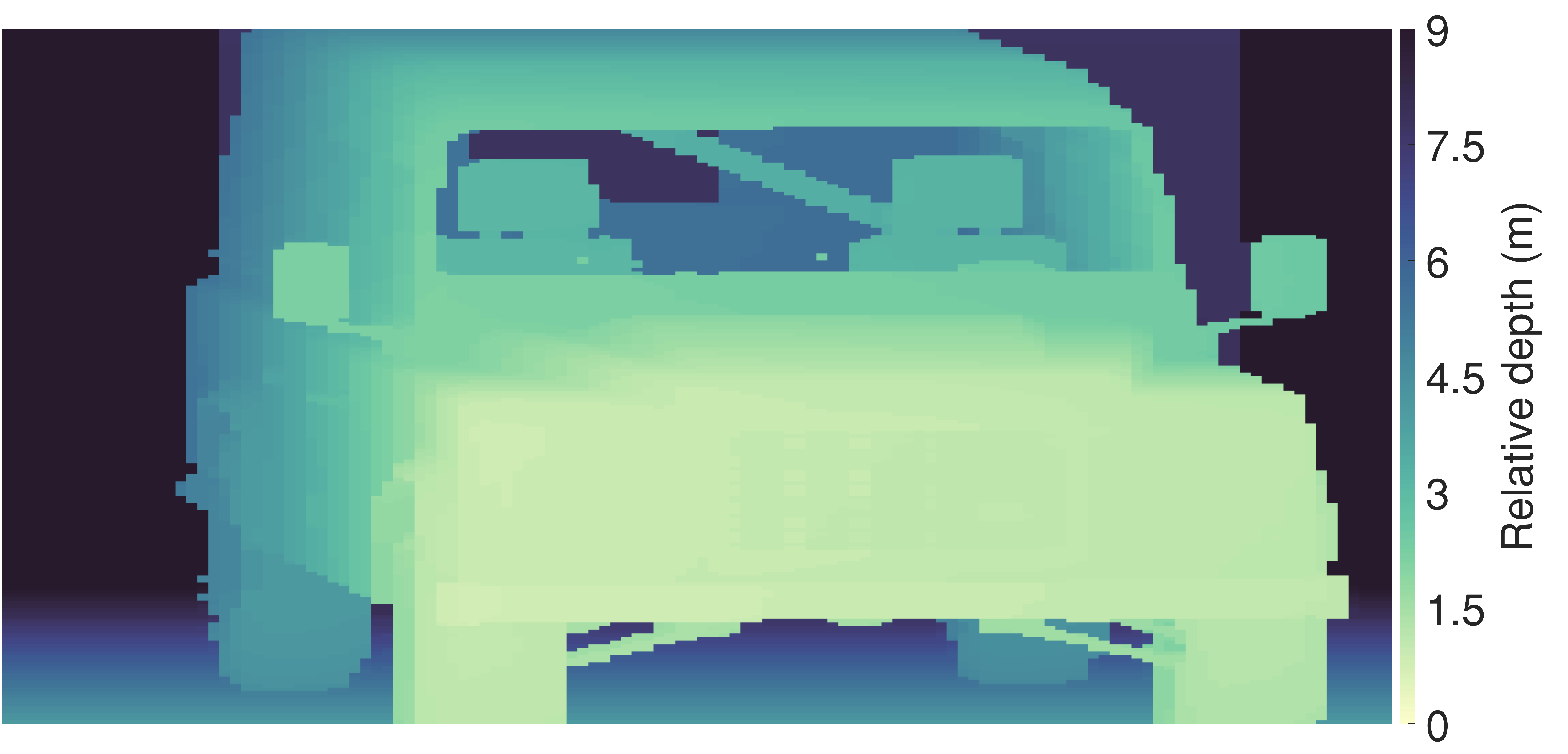}\\
        \bottomrule
        \end{tabular}
        \captionof{figure}{The reflectivity map (left) and ground truth depth map (right) created by the Unreal Engines' virtual environment. The reflectivity map accounts for both the reflectivity coefficient of the surface as well as its orientation relative to the camera (the surface normal). Note that the colorbar of the reflectivity map has been clamped to aid in visualization.}
	    \label{fig: Unreal_Ref}
\end{figure}

\begin{figure}[t!]
    %\begin{adjustbox}{max width=\columnwidth,center}
	\centering
	    \setlength\tabcolsep{2pt}
	    \footnotesize
        \begin{tabular}{r r r}
        \toprule
            &\multicolumn{1}{c}{\textbf{Simulation}}&\multicolumn{1}{c}{\textbf{Experiment}}\\
            \cmidrule(lr){2-2}
            \cmidrule(lr){3-3}
            \begin{turn}{90}\hspace{0.5cm}\textbf{Images}\end{turn}&\includegraphics[width=0.45\columnwidth]{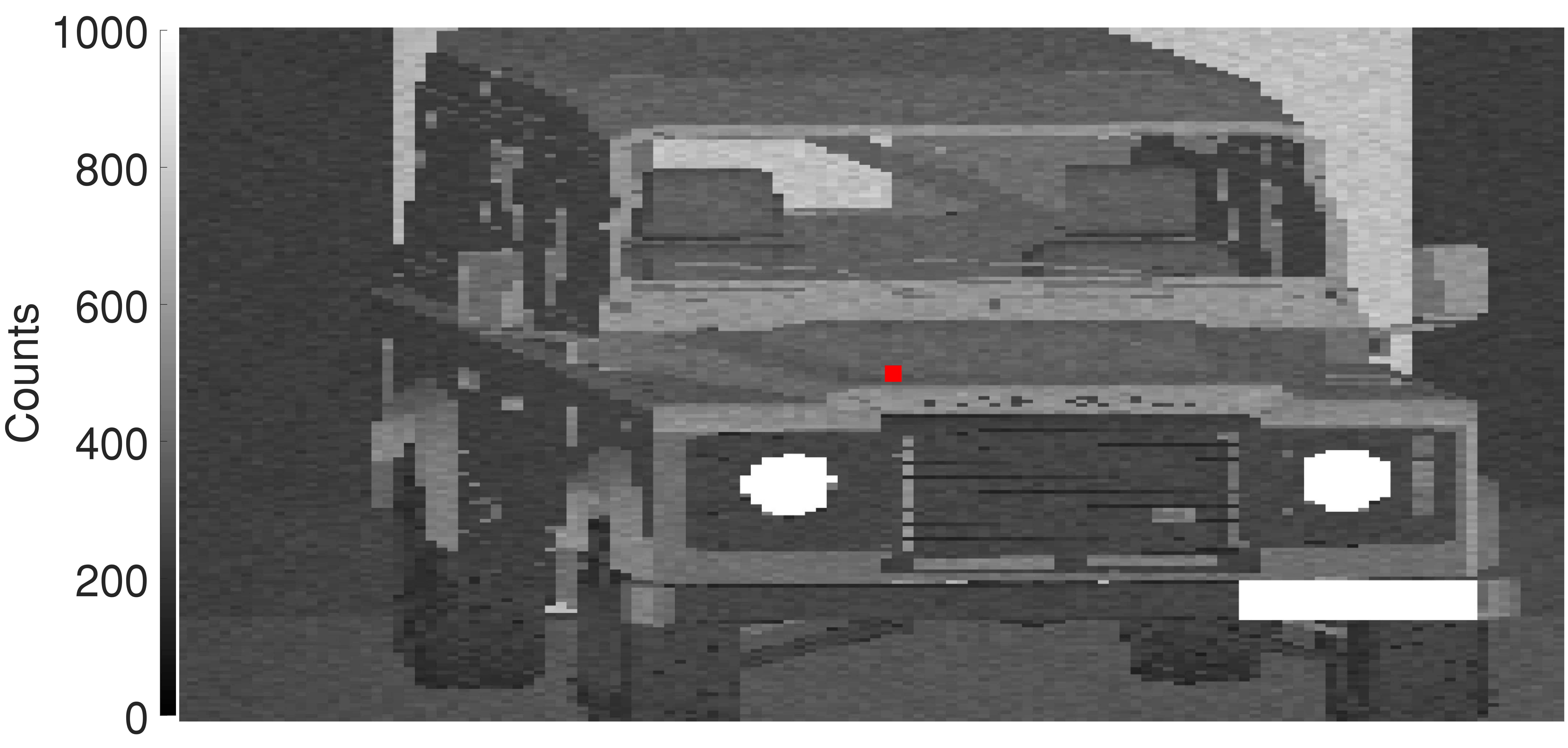}&\includegraphics[width=0.45\columnwidth]{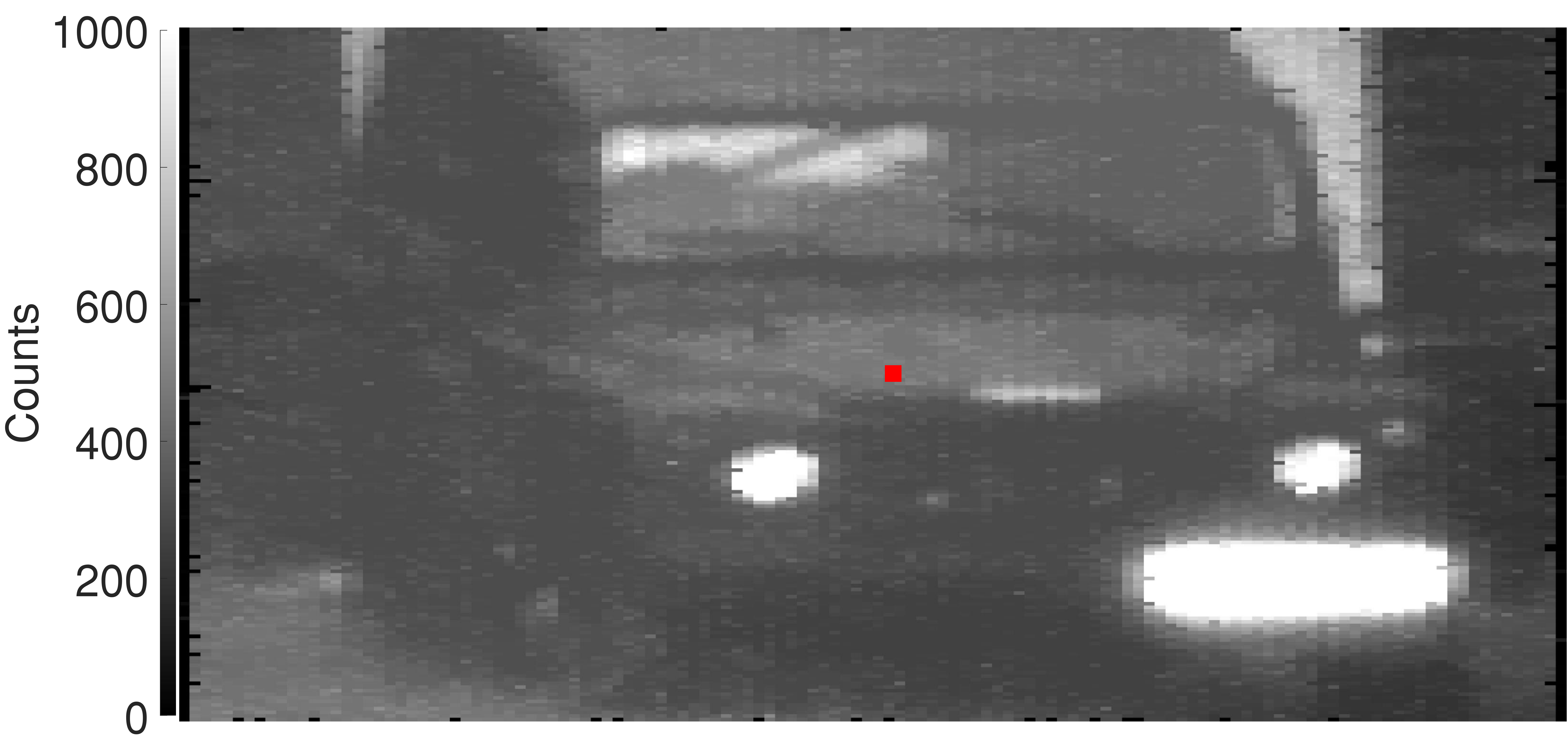}\\
            \cmidrule(lr){2-2}
            \cmidrule(lr){3-3}
            \begin{turn}{90}\hspace{0.5cm}\textbf{Histograms}\end{turn}&\includegraphics[width=0.43\columnwidth]{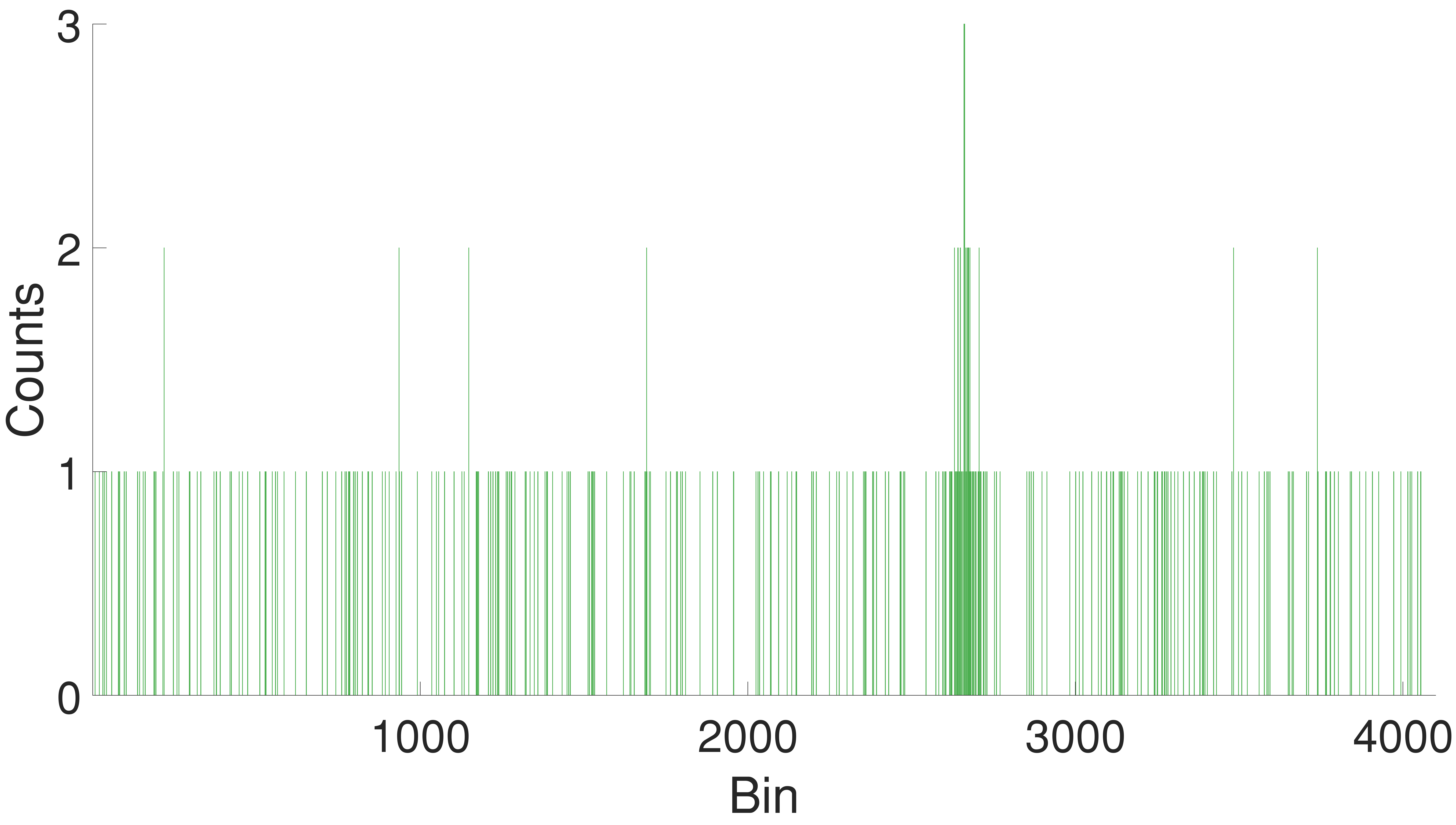}&\includegraphics[width=0.43\columnwidth]{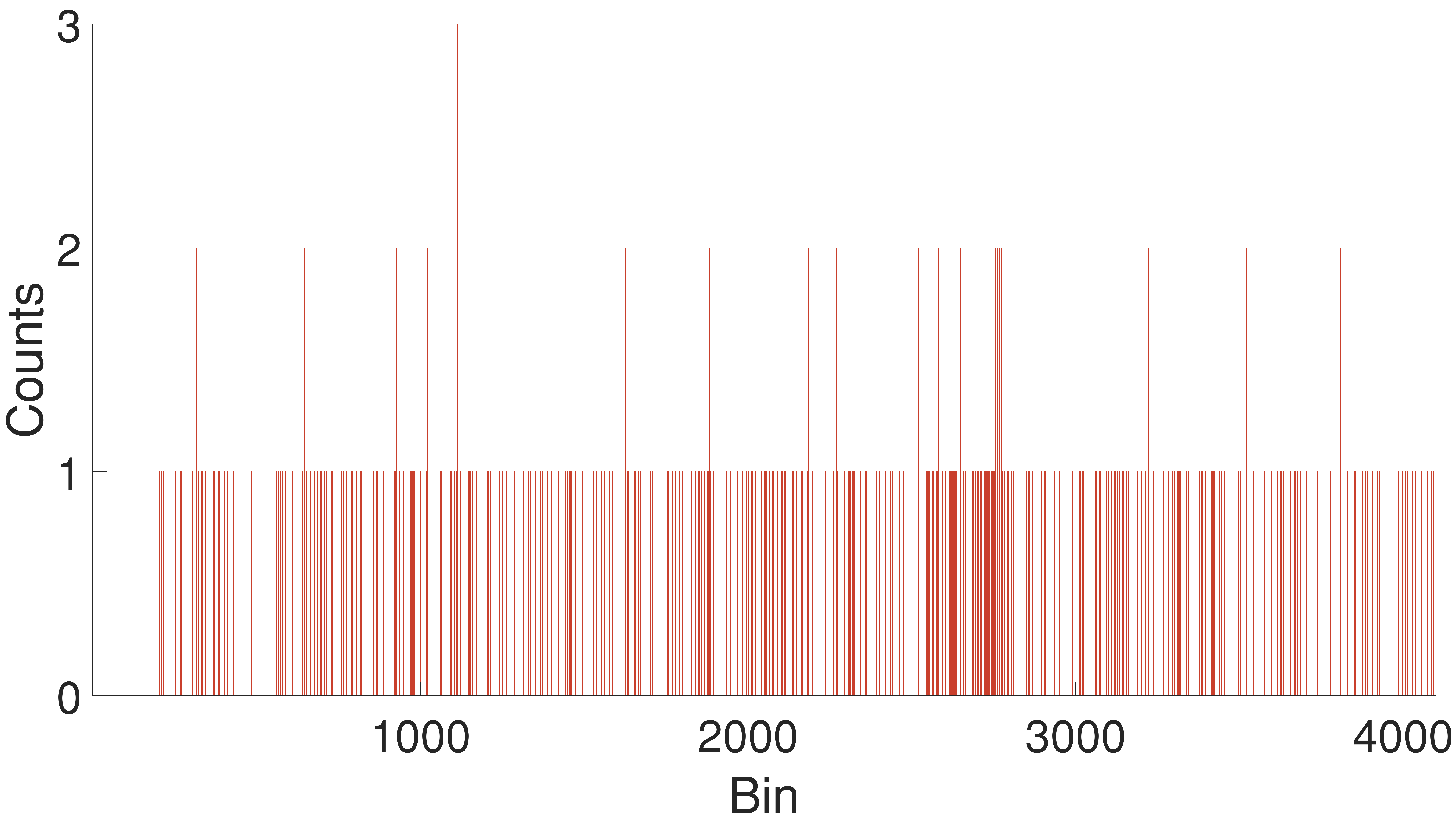}\\
            \cmidrule(lr){2-2}
            \cmidrule(lr){3-3}
            \begin{turn}{90}\hspace{0.75cm}\textbf{Intensities}\end{turn}&\includegraphics[width=0.44\columnwidth]{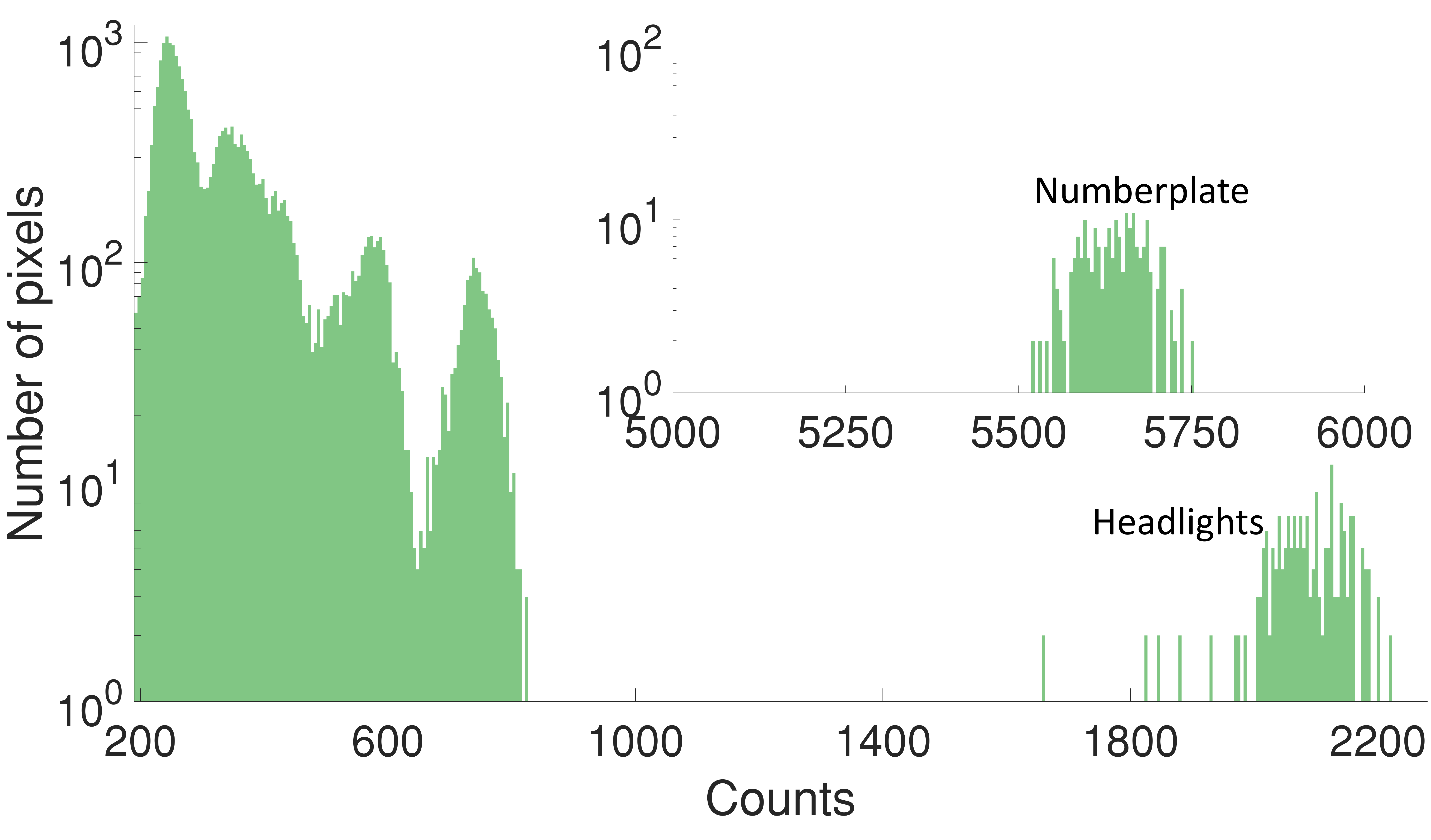}&\includegraphics[width=0.44\columnwidth]{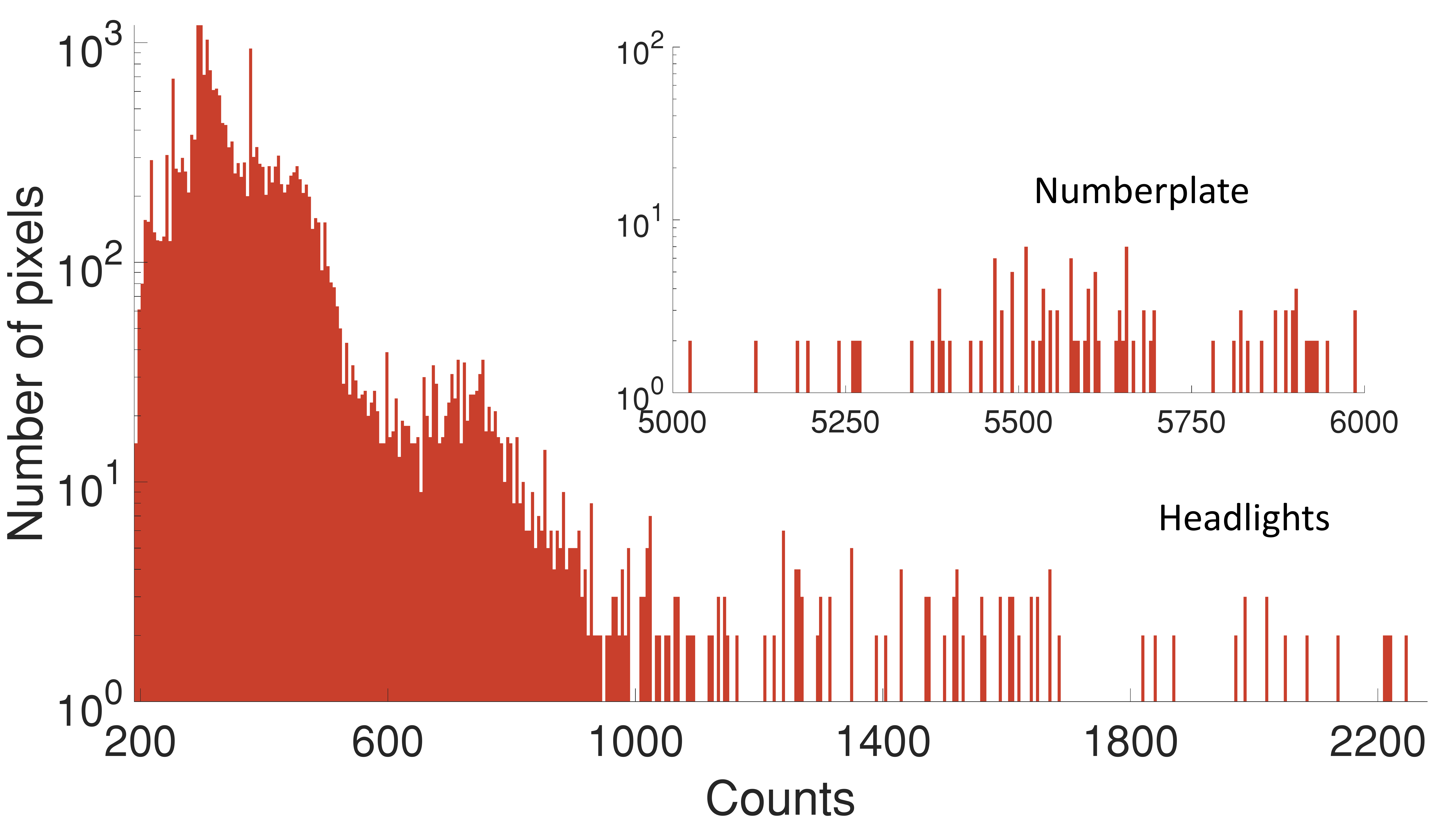}\\
        \bottomrule
        \end{tabular}
        \captionof{figure}{Quantitative and qualitative comparison between the simulated and measured intensity images for the Landrover. Top row; the simulated (left) and measured (right) intensity images. Middle row; the simulated (left) and measured (right) histograms for the pixel marked in red in the top row. Bottom row; the number of pixels (on a log scale) at a given intensity as a histogram for the simulated (left) and measured (right) results. The insets show the distribution at the high end of the counts (intensity) axis. Further, regions of notable intensity i.e., the numberplate and headlights have been annotated. }
	    \label{fig: Landrover_SPC}
\end{figure}

To illustrate the versatility of our approach the parameters in the model were adapted to reproduce a SPAD image of a Landrover\textsuperscript{\texttrademark} Wolf 110 parked in front of a wall captured at a range of 1.4 km. The data was captured under daylight conditions using a ``FLIMera" camera by Horiba\textsuperscript{\texttrademark}. This camera uses the same sensor as that examined in prior sections. A Cassegrain telescope was used to achieve the required focal length. For the full list of model parameters the reader is referred to Sec.~\ref{subsec:Experimental parameters used in simulating the landrover}. The experimental scene was recreated virtually within the Unreal Engine\textsuperscript{\texttrademark} using a 3D vehicle model which approximated the Landrover. The Unreal environment was used to create the reflectivity and depth maps shown in Fig.~\ref{fig: Unreal_Ref}. The reflectivity map accounts for both the reflectivity coefficients (assuming lambertian scattering) of the surfaces, which is contingent on the physical material of the surface, as well as the surface normals i.e., orientation of each surface relative to the camera. The depth map is used as the ground truth depth ($\mu$ in Eq.~\ref{eqn: histo prob img}) in the simulation.\\
\\
Figure~\ref{fig: Landrover_SPC} shows the simulated and measured intensity images of the Landrover. The simulation was performed using the model in histogram mode so as to match the SPAD data as closely as possible. The middle row of Fig.~\ref{fig: Landrover_SPC} compares, for a single pixel, the histogram generated by the model to the measured histogram. Additionally, Fig.~\ref{fig: Landrover_SPC} provides histograms showing the number of pixels at a given intensity. The qualitative and quantitative agreement between the images and histograms in Fig.~\ref{fig: Landrover_SPC} illustrates the models ability to produce reasonable approximations of SPAD images under real world conditions. Specifically, the ability to estimate the total photon return from specific components, in this instance the numberplate and headlights, as annotated in Fig.~\ref{fig: Landrover_SPC} on the scale of entire images is beneficial to the design process of imaging systems. Additionally, the ability to illustrate target or scenario specific features, such as the diagonal cross bar behind the drivers seat, and the wall visible through the rear window of the vehicle are beneficial in examining the performance of potential imaging systems. Note that the directions of the diagonal bars are reversed between the simulated and experimental image due to the drivers seat being on opposites sides between the 3D model and the measured vehicle. These differences in the 3D model, the assumption of Lambertian scattering, and the estimation of the reflectivity coefficients for the surfaces are the primary sources of the difference between the distribution of intensity histograms in Fig.~\ref{fig: Landrover_SPC}.\\
\begin{figure}[t!]
    %\begin{adjustbox}{max width=\columnwidth,center}
	\centering
	    \setlength\tabcolsep{2pt}
	    \footnotesize
        \begin{tabular}{r r r}
            \toprule
            &\multicolumn{1}{c}{\textbf{Simulation}}&\multicolumn{1}{c}{\textbf{Experiment}}\\
            \cmidrule(lr){2-2}
            \cmidrule(lr){3-3}
            \begin{turn}{90}\hspace{0.5cm}\textbf{Images}\end{turn}&\includegraphics[width=0.45\columnwidth]{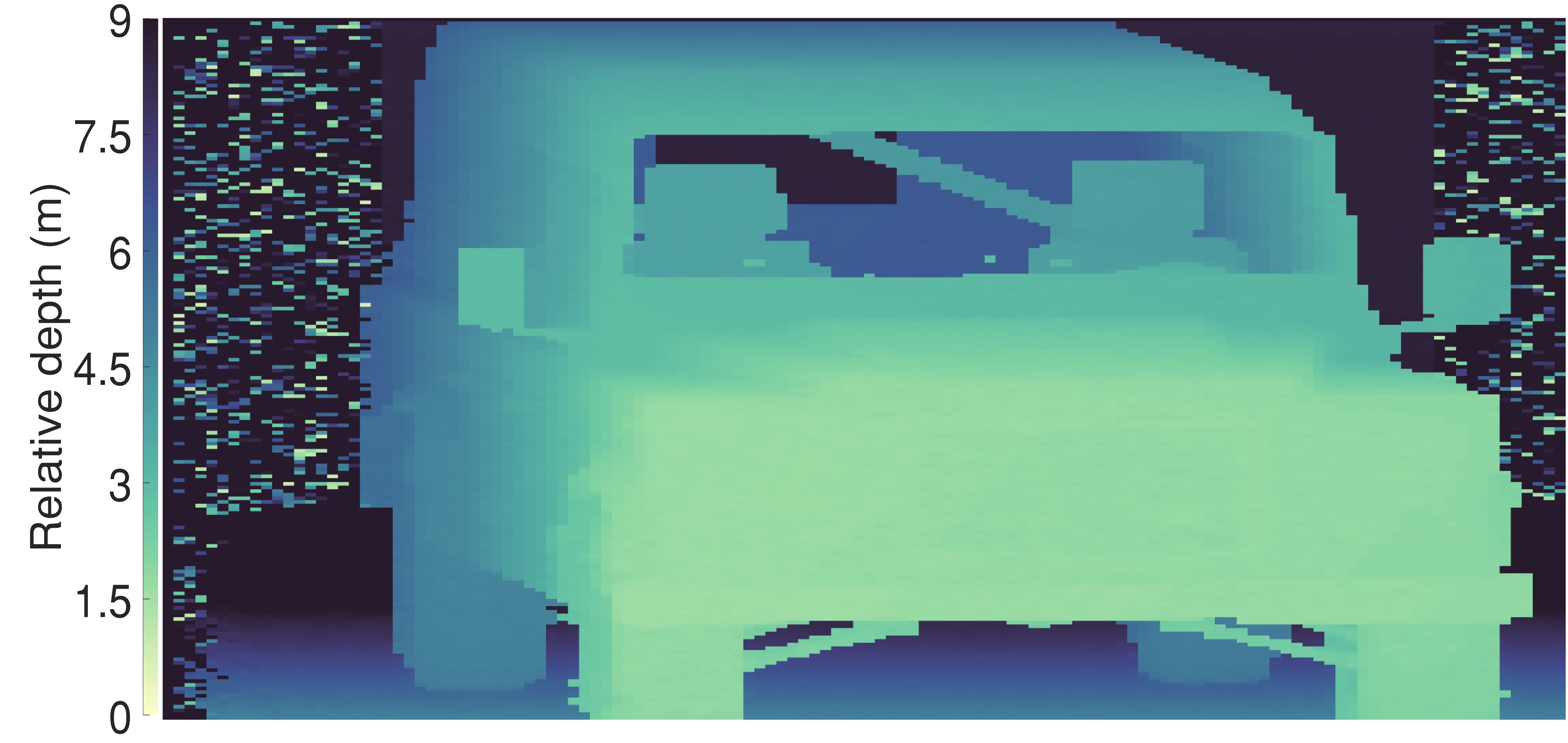}&\includegraphics[width=0.45\columnwidth]{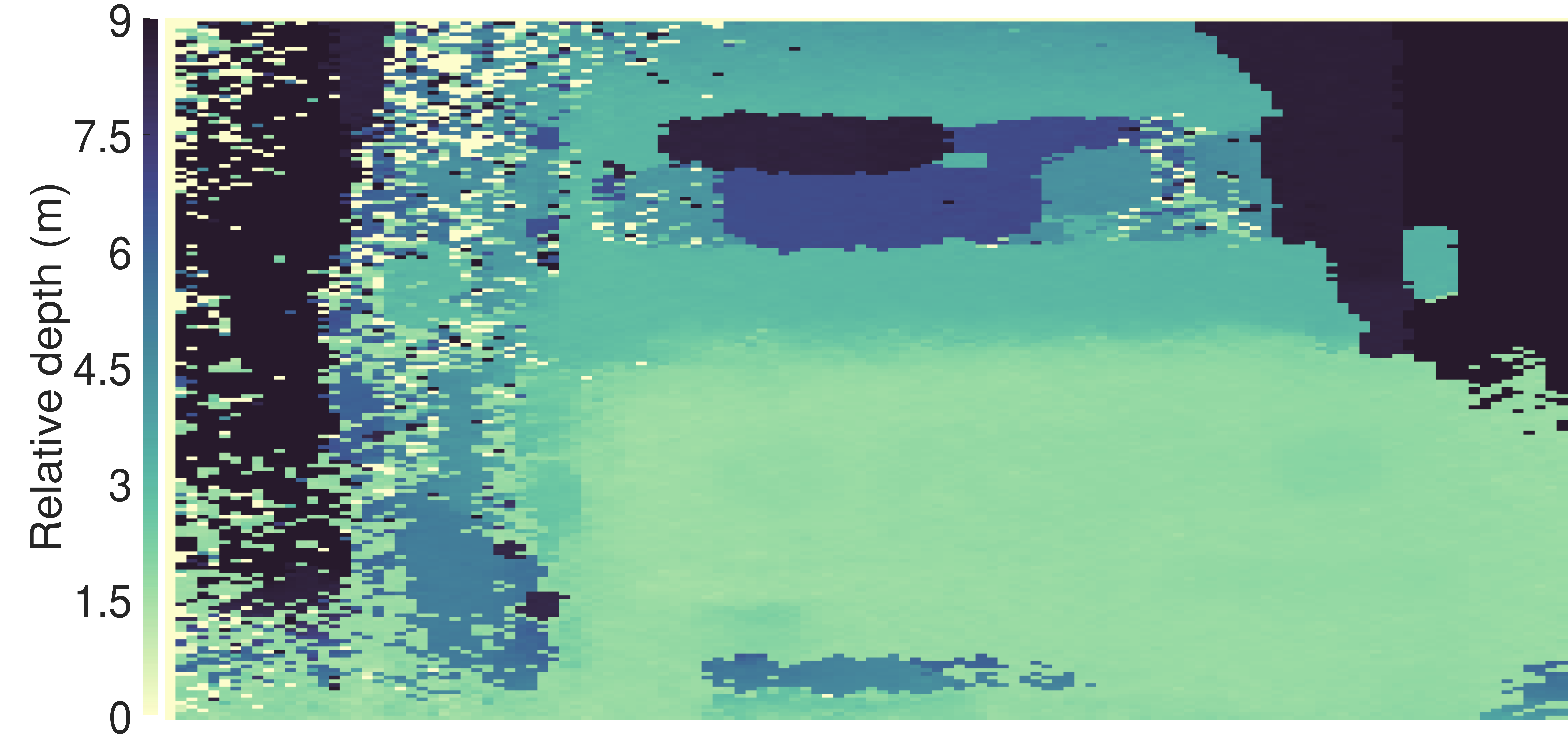}\\
            \cmidrule(lr){2-2}
            \cmidrule(lr){3-3}
            \begin{turn}{90}\hspace{0.9cm}\textbf{Depths}\end{turn}&\includegraphics[width=0.45\columnwidth]{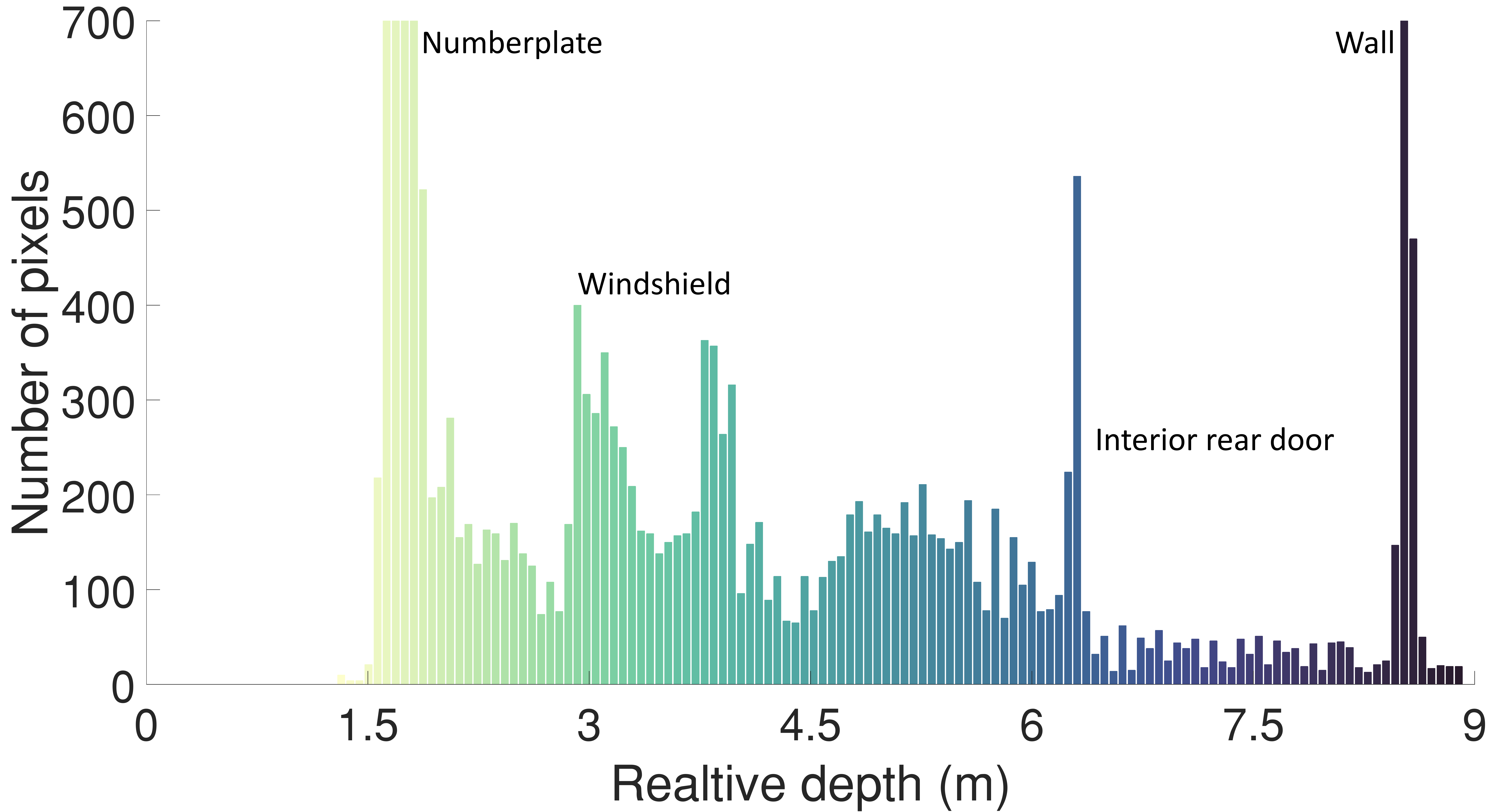}&\includegraphics[width=0.45\columnwidth]{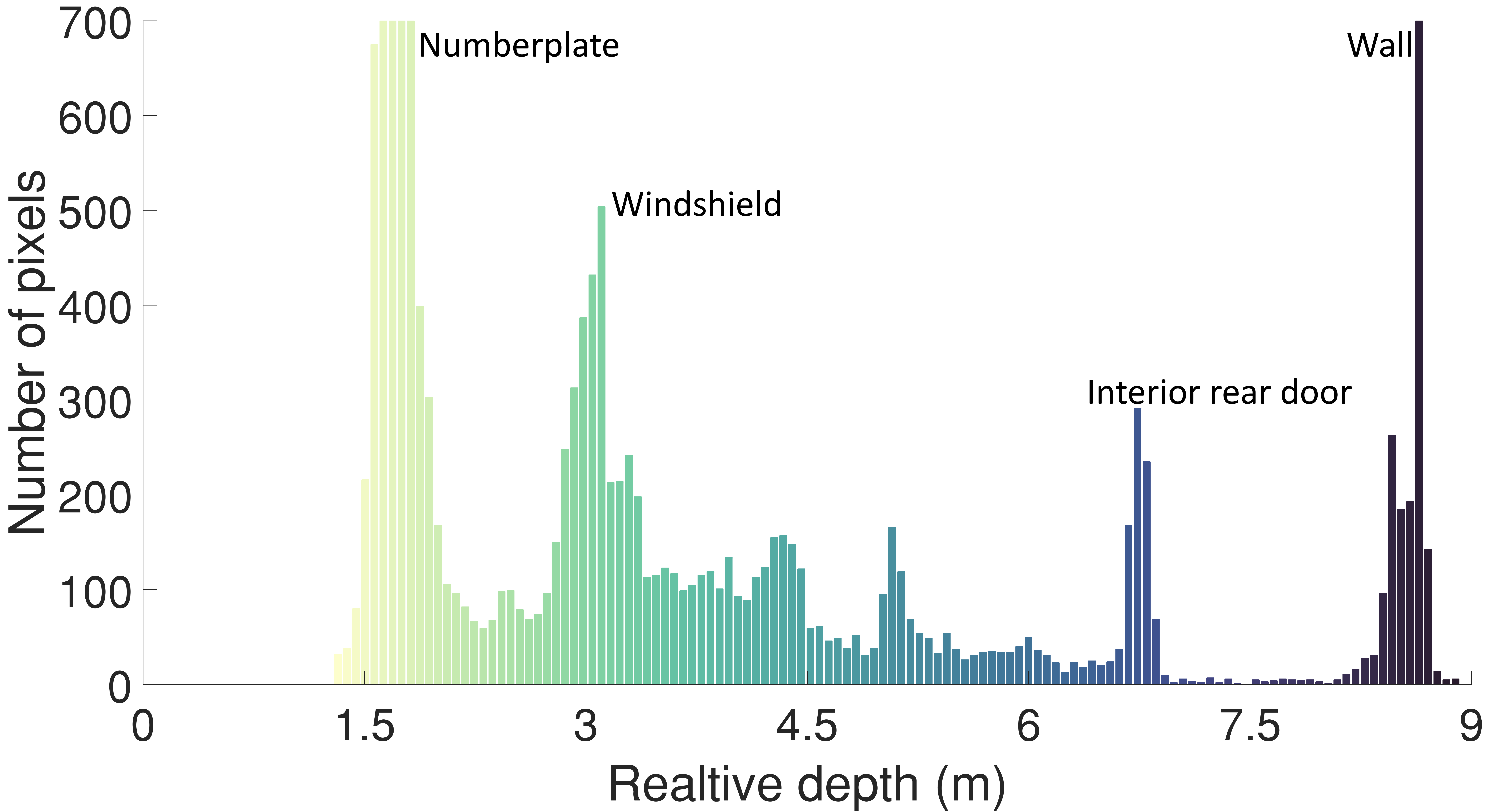}\\
        \bottomrule
        \end{tabular}
        \captionof{figure}{Quantitative and qualitative comparison between the simulated and measured depth images for the Landrover. Top row; the depth images derived from the simulated (left) and experimental (right) histograms respectively. Bottom row; the number of pixels at a given depth as a histogram for the simulated (left) and measured (right) results. The annotations represent the physical component in the image corresponding to the peak in the histogram.}
	    \label{fig: Landrover_TCSPC}
\end{figure}\\
\noindent Figure~\ref{fig: Landrover_TCSPC} shows the simulated and measured depth images of the Landrover. Additionally, Fig.~\ref{fig: Landrover_TCSPC} provides histograms showing the number of pixels at a given depth. As with Fig.~\ref{fig: Landrover_SPC} the ability to produce reasonable approximations of depth images together with the ability to attribute specific areas of a depth image to physical components (as annotated in the histograms of Fig.~\ref{fig: Landrover_TCSPC}) is beneficial when selecting or designing components for real world SPAD imaging systems. Notable in the experimental depth image in Fig.~\ref{fig: Landrover_TCSPC} is the region of constant depth in the lower-right corner of the image. This region can be attributed to the disproportionately high reflectivity of the numberplate (as seen in Fig.~\ref{fig: Unreal_Ref}  and~\ref{fig: Landrover_SPC}). This high reflectivity results in pixel cross talk as well as lens flaring effects, phenomena which are currently not accounted for by the model.\\
\\
Finally, we stress that the approach presented here, i.e., first determining the likelyhood of photon arrivals (Eq.~\ref{eqn: Likelyhood}) and then using a sampling routine (Eq.~\ref{eqn: histo prob img}), is applicable to a multitude of scenarios. For instance, to adapt the model for underwater imaging (or any environment in which back-scatter is significant) only Eq.~\ref{eqn: Likelyhood} needs to be changed. Specifically, by introducing a non-uniform back-scatter term (analogous to $P_{pp}$) the exponentially decaying background characteristic of scattering media can be modelled~\cite{tontini2020numerical,tobin2021robust}. Alternatively, by introducing multiple $P_{pp}$ terms to Eq.~\ref{eqn: Likelyhood} one could model multiple scattering returns per-pixel. These multiple returns are characteristic of the histograms obtained when using low transverse resolution integrating sensors, or imaging around corners~\cite{gariepy2016detection,hutchings2019reconfigurable}.
\section{Conclusion}
We have presented a tool for the accurate simulation of the images produced by SPAD based Lidar systems. We integrate a physical model of photon arrivals with a 3D virtual environment to create a physically significant description for the performance of SPAD imaging systems in the context of Fisher information. We present a computational model capable of either; leveraging the Cram\'er-Rao bound to rapidly produce large numbers of physically realistic SPAD images; or, operating in a regime where it produces realistic SPAD histograms on a pixel-by-pixel basis over the entire sensor. In this later regime we implement a Binomial sampling procedure which extends the simulation to the single-photon, single-bin limit whilst remaining consistent with prior Poisson based works. We comment on the effects that specific optical components have regarding the imaging abilities of SPAD systems and demonstrate the validity of our predictions with experimental measurements of a resolution test target. Finally, we illustrate the versatility of our approach by exploiting the 3D environment to correctly account for the complex surface geometries of a real vehicle. In so doing we present a complete image generation chain capable of accurately modelling the performance of SPAD systems under real world long range conditions. We expect this work to serve as a valuable tool in the ongoing development of SPAD enabled Lidar systems.  

\section*{Acknowledgements}
The authors gratefully acknowledge the funding support of the Defence Science Technologies Laboratory through project Dstlx-1000147352. This work was supported by EPSRC through grants EP/T00097X/1 and EP/S026428/1.\\
\\
The authors thank Dr. Abderrahim Halimi for his useful discussion on the nature of unbiased estimators.

\section{Appendices}
\subsection{Derivation of  photons detected per-pulse-per-pixel}
\label{subsec:Derivation of  photons detected per-pulse-per-pixel}
The number of photons returned from a target is described by a photon channel which models the loss of signal photons as a series of sequential processes. Consider a laser pulse of initial energy $E_0$ having a divergence $\theta$ projected over a range $R$, through an atmosphere of attenuation length $C_{atm}$. The energy density $\rho_E$ at the target is,
\begin{equation}
\begin{aligned}
    \rho_E = \frac{E_0e^{\frac{-R}{C_{atm}}}}{\pi R^2\tan^2(\theta)}.
\end{aligned}
\label{eqn: energy density}
\end{equation}
For an imaging sensor with pixels of effective size $W_p\times H_p$ at the focal plane of a collecting lens of focal length $f$ and f-number $f_{no}$, the energy $E_1$ available to each pixel is,
\begin{equation}
\begin{aligned}
    E_1 = \rho_E\left(\frac{R^2W_pH_p}{f^2}\right).
\end{aligned}
\label{eqn: pixel illum area}
\end{equation}
Assuming Lambertian reflection, for a target of reflectively $\Gamma$ the scattered energy $E_2$ which arrives at the aperture of the lens for each pixel is,
\begin{equation}
\begin{aligned}
    E_2 = \frac{\Gamma E_1e^{\frac{-R}{C_{atm}}}}{2\pi R^2}.
\end{aligned}
\label{eqn: scattered energy}
\end{equation}
Each pixel then captures a fraction of this scattered energy according to the aperture of the lens and the quantum efficiency $q$ of the detector,
\begin{equation}
\begin{aligned}
    E_3 = qE_2\pi\left(\frac{f}{2f_{no}}\right)^2.
\end{aligned}
\label{eqn: final energy}
\end{equation}
Combining Eqs.~\ref{eqn: energy density}-\ref{eqn: final energy} and dividing by $\frac{h c}{\lambda}$ where $\lambda$ is the wavelength of the illuminating light, the number of photons-per-pulse $P_{pp}$ captured by the detector is,
\begin{equation}
\begin{aligned}
    P_{pp} = \frac{\lambda E_0}{h c}\frac{q\Gamma e^{\frac{-2R}{C_{atm}}}}{8}\frac{ W_pH_p}{f_{no}^2\pi R^2\tan^2(\theta)}.
\end{aligned}
\label{eqn: photons per pulse sup}
\end{equation}
\subsection{Experimental parameters used in simulating the resolution test target}
\label{subsec:Experimental parameters used in simulating the resolution test target}
See Table~\ref{tab:full_params_1}
\begin{table}[h!]
    \centering
    %\footnotesize
        \begin{tabular}{c c c c}
	        \textbf{Symbol}&\textbf{Parameter}&\textbf{Value}&\textbf{Unit}\\
	        \cmidrule(lr){1-1}
	        \cmidrule(lr){2-2}
	        \cmidrule(lr){3-3}
	        \cmidrule(lr){4-4}
	        $E_0$&Energy per pulse&1&nJ\\
	        $\nu$&Repetition rate&2.25&MHz\\
	        $\lambda$&Wavelength&671& nm\\
	        $\sigma'$&Pulse FWHM&600&ps\\
	        \cmidrule(lr){1-4}
	        $R$&Range&14.73&m\\
	        $C_{atm}$&Attenuation&6.2&km\\
	        $\theta$&Divergence&0.02&radians\\
            \cmidrule(lr){1-4}
            $\Gamma$&Reflectivity&0.09&$-$\\
            $C_{bckg}$&Solar background&0&W\\
            \cmidrule(lr){1-4}
            $f_{no}$&f-number &2.0& -\\
            $C_{dc}$&Dark counts &126& Hz\\
            $\eta$&Exposure time&1000&$\mu$s\\
            $q$&Quantum efficiency&0.26&$-$\\
            $W_p/H_p$&Pixel size (width/height)&9.2&$\mu$m\\
            $\omega$&Bin width&50&ps\\
            $j$&Jitter&200&ps\\
        \bottomrule
        \end{tabular}
    \caption{The parameters used to model the resolution test target. Note that the target was effectively normal to all pixels and so a constant reflectivity value was used.}
    \label{tab:full_params_1}
\end{table}
\subsection{Match filtering as an unbiased estimator}
\label{subsec:Match filtering as an unbiased estimator}
Consider a single Gaussian signal on top of a uniform background $g(t)$ match filtered with a Gaussian kernel $f(t)$. The depth estimate $\hat{\mu}$ is then given by,
\begin{equation}
\begin{aligned}
    \hat{\mu} &= \text{argmax}[f(t)* g(t)],\\
\end{aligned}
\label{eqn: argmax}
\end{equation}
where
\begin{equation}
\begin{aligned}
    f(t) &= \frac{1}{\sigma'\sqrt{2\pi}}\exp{\left[-\frac{1}{2}\left(\frac{t-\mu}{\sigma'}\right)^2\right]},\\
    g(t) &= \frac{1}{\sigma'\sqrt{2\pi}}\exp{\left[-\frac{1}{2}\left(\frac{t-\mu'}{\sigma'}\right)^2\right]}+A\text{rect}\left[\frac{t-\mu'}{T}\right],\\
\end{aligned}
\label{eqn: f and g}
\end{equation}
Formally, the background is treated as a uniform value of amplitude $A$ over the time domain $T$ centered upon the true depth value $\mu'$. Under the conditions that the signal $g(t)$ contains only a single peak that is situated sufficiently far (i.e. >> 3$\sigma'$) from the boundaries of the domain of $t$, then Eq.~\ref{eqn: argmax} is equivalent to,
\begin{equation}
\begin{aligned}
    \max[f(t)*g(t)].\\
\end{aligned}
\label{eqn: max}
\end{equation}
Thus,
\begin{equation}
\begin{aligned}
    \max[f(t)*g(t)] &= \max\left[\int_{-\infty}^{\infty}f(t)g(t)dt\right],\\
    \max[f(t)*g(t)] &= \max\left\{B\exp{\left[-\frac{1}{2}\left(\frac{\mu-\mu'}{\sigma'}\right)^2\right]}+\right.\\
    &\left.\int_{-\infty}^{\infty}f(t)\text{rect}\left[\frac{t-\mu'}{T}\right]dt\right\}.
\end{aligned}
\label{eqn: inte}
\end{equation}
Here $B$ is a normalization constant. Equation~\ref{eqn: inte} implies that under the prior stated conditions,
\begin{equation}
\begin{aligned}
    \int_{-\infty}^{\infty}f(t)\text{rect}\left[\frac{t-\mu'}{T}\right]dt& = C\\
    &\implies \max[f(t)*g(t)]\iff \mu = \mu'\\
    &\implies \hat{\mu} = \mu = \mu'
\end{aligned}
\label{eqn: mu cond}
\end{equation}
where $C$ is an integration constant. Hence, Eq.~\ref{eqn: mu cond} implies that match filtering by a Gaussian kernel is an unbaised estimator for the case of a single Gaussian signal on a uniform background.\\
\\
Further, in the case of a Gaussian signal, in the absence of background ($A=0$), the minimum variance unbiased estimator (i.e. the estimator which saturates the Cram\'er-Rao bound) is given by the maximum likelyhood~\cite{cramer1946mathematical}. This operation is equivalent to Eq.~\ref{eqn: argmax} with a kernel $h(t) = \log[f(t)]$. However, for $A\neq0$, $h(t)$ is biased towards the center of the domain $T$ and as such is no longer consistent with the conditions for the Cram\'er-Rao bound. Consider the kernel $f(t)$ then,
\begin{equation}
\begin{aligned}
    \log[A+f(t)]&= \log(A)+\frac{f(t)}{A}+\textit{o}\left(\frac{f(t)}{A}\right)\\
\end{aligned}
\label{eqn: log A}
\end{equation}
Here $\textit{o}()$ represents the higher order terms in the Taylor series. Hence, as $A$ (i.e. the amplitude of background) increase, the contribution of these terms diminishes. Hence   
\begin{equation}
\begin{aligned}
    \log[A+f(t)]\propto D+\frac{f(t)}{A}\\
\end{aligned}
\label{eqn: kernel approx}
\end{equation}
Specifically, $D$ is a constant only tied to the amplitude of the background while the form of the kernel $f(t)$ remains unchanged. Consequently, $f(t)$ represents an operation analogous to the maximum likelyhood for cases with $A\neq0$. 
\subsection{Experimental parameters used in simulating the landrover}
\label{subsec:Experimental parameters used in simulating the landrover}
\begin{table}[h!]
    \centering
    %\footnotesize
        \begin{tabular}{c c c c}
	        \textbf{Symbol}&\textbf{Parameter}&\textbf{Value}&\textbf{Unit}\\
	        \cmidrule(lr){1-1}
	        \cmidrule(lr){2-2}
	        \cmidrule(lr){3-3}
	        \cmidrule(lr){4-4}
	        $E_0$&Energy per pulse&14&$\mu$J\\
	        $\nu$&Repetition rate&33&kHz\\
	        $\lambda$&Wavelength&532& nm\\
	        $\sigma'$&Pulse FWHM&3.5&ns\\
	        \cmidrule(lr){1-4}
	        $R$&Range&1.4&km\\
	        $C_{atm}$&Attenuation&6.2&km\\
	        $\theta$&Divergence&1.07&milliradian\\
            \cmidrule(lr){1-4}
            $\Gamma$&Body reflectivity&0.065&$-$\\
            &Tyres and trim reflectivity&0.029&$-$\\
            &Wall reflectivity&0.081&$-$\\
            &Ground reflectivity&0.066&$-$\\
            &Seats reflectivity&0.04&$-$\\
            &Heatlights reflectivity&0.25&$-$\\
            &Numberplate reflectivity&0.8&$-$\\
            $C_{bckg}$&Solar background&0.5&W\\
            \cmidrule(lr){1-4}
            $f_{no}$&f-number &10.0& -\\
            $C_{dc}$&Dark counts &126& Hz\\
            $\eta$&Exposure time&83&$\mu$s\\
            $q$&Quantum efficiency&0.26&$-$\\
            $W_p/H_p$&Pixel size&9.2&$\mu$m\\
            $\omega$&Bin width&50&ps\\
            $j$&Jitter&1.5&ns\\
        \bottomrule
        \end{tabular}
    \caption{The parameters used to model the Landrover. Note that the reflectivities represent the base values prior to their modification based on the orientation of the surface relative to the camera.}
    \label{tab:full_params_2}
\end{table}
% Bibliography
\bibliographystyle{ieeetr}
\bibliography{References}

% Full bibliography added automatically for Optics Letters submissions; the following line will simply be ignored if submitting to other journals.
% Note that this extra page will not count against page length

%Manual citation list
%\begin{thebibliography}{1}
%\bibitem{Zhang:14}
%Y.~Zhang, S.~Qiao, L.~Sun, Q.~W. Shi, W.~Huang, %L.~Li, and Z.~Yang,
 % \enquote{Photoinduced active terahertz metamaterials with nanostructured
  %vanadium dioxide film deposited by sol-gel method,} Opt. Express \textbf{22},
  %11070--11078 (2014).
%\end{thebibliography}

\end{document}